\documentclass[twocolumn,prl,superscriptaddress,bibnotes,notitlepage,nofootinbib]{revtex4-1}
\usepackage{color}
\definecolor{red}{rgb}{0.75,0,0}
\definecolor{blue}{rgb}{0,0,0.75}
\definecolor{green}{rgb}{0,0.5,0}
\newcommand{\SK}[1]{\textcolor{black}{#1}} 

\usepackage{amsmath}
\usepackage{amssymb}
\usepackage{lipsum}
\usepackage{bm}
\usepackage{xcolor}

\usepackage{graphicx}
\usepackage{verbatim}
\usepackage[T1]{fontenc}
\usepackage{hyperref}
\usepackage{mathrsfs}  
\usepackage[T1]{fontenc}

\def\be{\begin{equation}}
\def\ee{\end{equation}}
\def\bea{\begin{eqnarray}}
\def\eea{\end{eqnarray}}

\def\D2{\boldsymbol{\mathcal{D}}}
\def\besub{\begin{subequations}}
\def\eesub{\end{subequations}}

\def\bwd{\begin{widetext}}
\def\ewd{\end{widetext}}

\definecolor{ao(english)}{rgb}{1.0, 0.0, 0.0}
\definecolor{armygreen}{rgb}{0.29, 0.33, 0.13}
\definecolor{auburn}{rgb}{0.43, 0.21, 0.1}
\definecolor{brightmaroon}{rgb}{0.76, 0.13, 0.28}
\definecolor{cadmiumred}{rgb}{0.89, 0.0, 0.13}
\definecolor{carnelian}{rgb}{0.7, 0.11, 0.11}
\definecolor{cornellred}{rgb}{0.7, 0.11, 0.11}
\definecolor{crimsonglory}{rgb}{0.75, 0.0, 0.2}
\definecolor{orangeyellow}{rgb}{0.3, 0.2, 0.2}
\definecolor{fluorescentorange}{rgb}{1.0, 0.75, 0.0}
\definecolor{gamboge}{rgb}{0.89, 0.61, 0.06}
\newcommand{\bsf}[1]{\textsf{\textbf{#1}}}

\newcommand{\AMNNN}[1]{\textcolor{black}{#1}}

\newcommand{\AMN}[1]{\textcolor{black}{#1}}

\newcommand{\SR}[1]{\textcolor{black}{#1}}

\newenvironment{frcseries}{\fontfamily{frc}\selectfont}{}

\newcommand{\AMF}[1]{\textcolor{black}{#1}}

\begin{document}
\title{Active cholesterics: odder than odd elasticity}
\author{S. J. Kole}
\email{swapnilkole@iisc.ac.in}
\affiliation{Centre for Condensed Matter Theory, Department of Physics, Indian Institute of Science, Bangalore 560 012, India}
\author{Gareth P. Alexander}
\email{G.P.Alexander@warwick.ac.uk}
\affiliation{Department of Physics and Centre for Complexity Science, University of Warwick, Coventry CV4 7AL, United Kingdom}
\author{Sriram Ramaswamy}
\email{sriram@iisc.ac.in}
\affiliation{Centre for Condensed Matter Theory, Department of Physics, Indian Institute of Science, Bangalore 560 012, India}
\author{Ananyo Maitra}
\email{nyomaitra07@gmail.com}
\affiliation{Sorbonne Universit\'{e} and CNRS, Laboratoire Jean Perrin, F-75005, Paris, France}

\begin{abstract}
In equilibrium liquid crystals, chirality leads to a variety of spectacular three-dimensional structures, but chiral and achiral phases with the same broken continuous symmetries have identical long-time, large-scale dynamics. In this paper, we demonstrate that chirality qualitatively modifies the dynamics of layered liquid crystals in active systems in both two and three dimensions due to an active ``odder'' elasticity. In three dimensions, we demonstrate that the hydrodynamics of active cholesterics differs fundamentally from smectic-A liquid crystals, unlike their equilibrium counterpart. This distinction can be used to engineer a columnar array of vortices, with anti-ferromagnetic vorticity alignment, that can be switched on and off by external strain. A two-dimensional chiral layered state -- an array of lines on an incompressible, free-standing film of chiral active fluid with a preferred normal direction -- is generically unstable. However, this instability can be tuned in easily realisable experimental settings, when the film is either on a substrate or in an ambient fluid. 
\end{abstract}

\maketitle

Chiral molecules form a spectacular range of liquid-crystalline phases \cite{TGB, Mermin_RMP,deGen} at thermal equilibrium, of which the best known is the cholesteric, with a helical structure in which the molecular orientation, described by a headless unit vector called the director $\hat{{\bf n}}$, spontaneously twists at a uniform rate $q_0$ along the pitch axis \cite{deGen}. This uniform periodic modulation does not break translational invariance: unlike in a density wave, all surfaces of constant phase are equivalent, and an arbitrary translation along the pitch axis can be compensated by a rotation about it. Nevertheless, at scales much larger than $1/q_0$, the mechanics of a cholesteric is precisely the same as that of a smectic A which has an achiral one-dimensional \emph{density} modulation. That is, microscopic chirality leads to a one-dimensional periodic structure, but the asymptotic long-wavelength elasticity and hydrodynamics of this structure show no signature of chirality \cite{Lub_chol, Rad_Lub,opticsfoot}. In this paper we show that this equivalence does not carry over to \emph{active} cholesteric and smectic A phases \cite{Tap_chol,Tap_smec}, thanks to effects that go beyond the ``odd elasticity'' of chiral active solids \cite{Deboo2}. 
\begin{figure}[b]
\label{Vortlat}
\centering
\includegraphics[width=\linewidth]{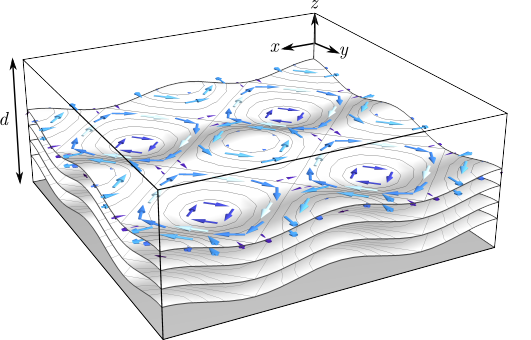}
\caption{The spontaneous vortex-lattice state: the chiral active force density (governed by $z_c$) generates counterrotating circulatory flows, with vorticity $\Omega_z$ obeying $\eta\nabla^2\Omega_z=-z_c\nabla^2\nabla_\perp^2 u$, around the undulations $u$ of the active Helfrich-Hurault instability.}
\label{fig:3dHH}
\end{figure}

Recall that active matter is matter with a sustained supply of free energy, and hence broken detailed balance, at the scale of its constituents. This microscale drive manifests itself macroscopically as nonequilibrium currents and forces \cite{SRJSTAT, LPDJSTAT, RMP, SRrev,Prost_nat, SalJul, TonTuRam, CatesRev, Curierep, Shelleyrev}. Continuum hydrodynamic theories of fluid \cite{scalar}, liquid-crystalline and crystalline phases \cite{TonTuRam, Prost_nat, Tap_smec, Tap_chol, smecp, Chen_Toner, Ano_sol} of active matter have been constructed including extensions with chiral asymmetry \cite{Lewis1, Lewis2, Strempel, Cates_drop, Deboo2, Seb1, Cates_chi, Carles, Luca_chi, Ano_chi}. In this paper, we construct theories of layered active chiral systems. Our hydrodynamic theory applies equally to cholesterics in the strict sense and to smectic A phases composed of homochiral units. We will refer to both as active cholesterics. We also construct the theory of an active two-dimensional (2D) chiral smectic, which could arise if three-dimensionally chiral particles were restricted to a thin film with a distinguished normal direction. 

\begin{figure}[t]
\centering
\includegraphics[width=\linewidth]{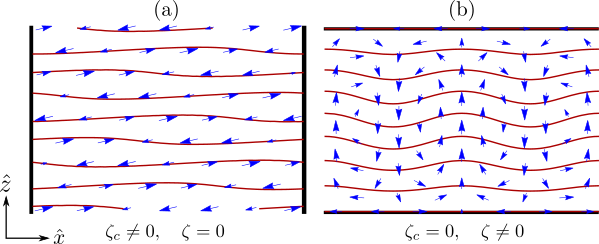}
\caption{Active instabilities in two-dimensional layered states. (a) Instability arising from the chiral active force ($\zeta_c$, here positive). The layers are indicated by dark red lines, while the linear flow field at instability overlaid (blue arrows). The thick black lines indicate confining walls, which create a finite threshold for the instability. (b) Instability arising from the achiral active force ($\zeta$) for comparison; same stylings.}
\label{fig:2d}
\end{figure}

Our central finding is that \AMF{active cholesterics possess a chiral stress corresponding to a non-existent component of the strain tensor that yields a force density tangent to contours of constant mean curvature of the layers.}  
As a result \AMF{of this odder than odd elasticity}, the undulational instability created by active stresses \cite{Tap_smec,Tap_chol} leads to spontaneous vortical flow arranged in a two-dimensional array with vorticity aligned along the pitch axis and alternating in sign in the plane (Fig.~\ref{fig:3dHH}). This vortex-lattice state can be switched on or off by means of an externally imposed uniaxial stress. Lastly, a two-dimensional active cholesteric is unstable with an activity threshold that goes to zero for an infinite system. This tilted-varicose instability (Fig.~\ref{fig:2d}) is however not inevitable, as we discuss later in the paper. 
 
We now show how we obtain these results. \AMF{A pattern-formation framework \cite{HohenCross} offers a foolproof approach to the construction of the hydrodynamic equations for active cholesterics, equivalent to the traditional route \cite{Tap_smec, Tap_chol} starting with the equations of motion for an orientation field and eliminating the fast degrees of freedom.} Accordingly, we begin by extending \cite{CatesH1, CatesH2} to define \textit{active model H*}: the coupled dynamics of a pseudoscalar density $\psi$ governed by a conservation law  $\partial_t\psi = - \nabla \cdot {\bf J}$ and a momentum density $\rho {\bf v}$ whose dynamics in the Stokesian regime is governed by $\nabla\cdot\bm{\sigma}=0$, with current ${\bf J} = \psi {\bf v} -M \nabla \mu + {\bf J}_a + {\bf J}_c$ and stress tensor $\boldsymbol{\sigma} =  -\eta [\nabla {\bf v} + (\nabla {\bf v})^T]  + \boldsymbol{\sigma}_{\psi} + {p} \bsf{I} - \boldsymbol{\sigma}_c  - \boldsymbol{\sigma}_a$, where subscripts $a$ and $c$ denote achiral active and chiral  contributions respectively. Here $M$ is a mobility, $\mu = \delta F/ \delta \psi$ is a chemical potential expressed in terms of a free-energy functional $F[\psi]$, $\eta$ is a viscosity, the passive force density $-\nabla \cdot \boldsymbol{\sigma}_{\psi} = -\psi \nabla \mu$ is the Onsager counterpart to $\psi {\bf v}$ \cite{deGroot}, and the \SK{pressure $p$} imposes overall incompressibility $\nabla \cdot {\bf v} = 0$. ${\bf J}_a=\lambda_{1} \psi \nabla \psi \nabla^2 \psi + \lambda_2 \psi \nabla (\nabla \psi)^2$ as familiar from active models B and H \cite{CatesH1, CatesH2, AMB1, AMB2, Cesare1, CesarePRX}. In what follows we ignore the chiral currents ${\bf J}_c$, whose effects on the dynamics of layered states arise at sub-leading order in wavenumber \cite{supp}. The achiral active stress \cite{CatesH1, CatesH2}, in both two and three dimensions, is $\boldsymbol{\sigma}_a=\zeta_H \nabla \psi \nabla \psi$ while the chiral active stress is 
\begin{align}
\label{stress2and3} 
(\sigma_{c})_{{ij}} = \AMN{\bar{z}_c} \partial_l(\epsilon_{ijk}\partial_k\psi\partial_l\psi), \,
\, &d=3;&\nonumber \\ 
\boldsymbol{\sigma}_c =
\bar{\zeta}_c\boldsymbol{\varepsilon}\cdot \nabla \psi \nabla \psi, \, &d=2,& 
\end{align}
where $d=2$ corresponds to a thin film of 3D chiral material with a distinguished normal taken to be along ${\bf N}\equiv +\hat{\bf y}$, thus inheriting uniquely the two-dimensional antisymmetric tensor $\boldsymbol{\varepsilon}$ {with components} $\varepsilon_{ij}=\epsilon_{ikj}N_{k}$. 
Though here written as an antisymmetric stress $(\sigma_{c})_{{ij}}$ can be given in an equivalent symmetric form and is allowed in momentum-conserving systems~\cite{Cates_chi,MPP,StarkLubensky,supp}. 

A Swift-Hohenberg free-energy functional $F$ \cite{SH, supp} allows model H$^{*}$ to describe the dynamics of spatially modulated states $\psi =\psi_0 + \psi_1$ where $\psi_1$, with zero spatial average, represents a modulation with wavelength $2 \pi /q_s$ about a uniform background $\psi_0$. We consider the dynamics about a steady state with a one-dimensional spatial modulation\AMF{, ${\psi}_1={\psi}_1^0(e^{i{\phi}}+e^{-i{\phi}})$, with $\phi={q}_s(z-u)$, describing a periodic array of parallel lines or planes of constant phase, in $d=2$ or $3$ respectively, with normal along $\hat{\bf z}$, with small fluctuations $u({\bf r},t)$. We begin with $d=3$. Defining the scaled phase-gradient vector, which is parallel to the normal of the fluctuating layers, as ${\bf n}=\nabla\phi/q_s=\hat{\bf z}-\nabla u(x,y,z,t)$ we obtain the dynamical equation of the displacement field of the layers from their mean positions \cite{supp}:}  
\begin{equation}
\label{ueqn3d}
\partial_t{u}={{\bf v}}\cdot{{\bf n}}+{\Lambda}_{1}{{\bf n}}\cdot\nabla{E}+{\Lambda}_{2}{\nabla}\cdot{{\bf n}}(1-2{E})-{\Gamma}_u\frac{\delta {F}[u]}{\delta {u}}, 
\end{equation}
where ${\Lambda}_{1}=-2\psi_1^{0^2}q_s^2(\lambda_1+\lambda_2)$ and ${\Lambda}_{2}=2\psi_1^{0^2}q_s^2\lambda_2$ are active, achiral permeative terms, the final term is passive permeation with $\Gamma_u=-Mq_s^2$, and  $E=\partial_z u-(1/2)(\nabla u)^2$ is the covariant strain. Finally, $F[u]=(1/2)\int [BE^2+K(\nabla^2 u)^2]$ is the rotation-invariant free energy \cite{deGen, GrinPel} which would have controlled the relaxational dynamics of the cholesteric state in the absence of activity with $B$ being the layer-compression modulus and $K$ being the bending rigidity of the layers which can be expressed in terms of the coefficients in the Swift-Hohenberg free energy \cite{supp}. Force balance for our system takes the form 
\begin{equation}
\label{vel3du}
\eta\nabla^2{v}_i=n_i\frac{\delta {F}[u]}{\delta {u}}+\partial_i p+\partial_j[{\zeta}w_{ij}+{z}_{c}\partial_l(\epsilon_{ijk}w_{kl})]
\end{equation}
where $w_{ij}$ encodes the active stresses from \eqref{stress2and3} and preceding. To linear order in displacements, with \AMF{$\perp$ denoting directions transverse to $\hat{{\bf z}}$,} $\bm{w}_{z\perp}=\bm{w}_{\perp z}=\nabla_\perp u$, $w_{zz}=2\partial_z u$ and all other components are $0$, $\zeta=\psi_1^{0^2}q_s^2\zeta_H$, $z_c=\psi_1^{0^2}q_s^2\bar{z}_c$ and the pressure $ p$ enforces three-dimensional incompressibility $\nabla\cdot{\bf v}=0$. 
\AMF{The term proportional to $z_c$ in \eqref{vel3du} is the chiral active force density. It is the curl of the vector $\partial_l w_{kl}$ and is therefore divergence-free. 
Expressed as a vector this chiral active force density is $-z_c \,\hat{\bf z} \times \nabla_{\perp}(\nabla_{\perp}^2 u + \partial_{zz} u)$ and is directed (primarily) tangentially to contours of constant mean curvature of the layer undulation, driving the vortical flow shown in Fig.~\ref{fig:3dHH}.} 
\AMF{Like odd elasticity, the $z_c$-term in \eqref{vel3du} is a parity-breaking stress in response to layer displacements. Unlike the odd elastic force density of two-dimensional chiral active solids \cite{Deboo2}, which arises from an antisymmetry in the linear relation between stress and strain, this cholesteric chiral force density \eqref{vel3du} arises even when the strain $E=0$.}

\AMF{It might seem that $z_c$ does not affect the hydrodynamics of the layered state as it appears at a higher order in gradients than the achiral active force. Indeed, it does not affect the linear dynamics of the displacement field at all: the eigenfrequency for displacement fluctuations to leading order in wavenumber, obtained by projecting \eqref{vel3du} transverse to the wavevector and solving for the velocity field yields, is} $\omega=-(i/\eta q^4)(Bq_z^2q_\perp^2-\zeta q_\perp^2q^2)$. Therefore, as noted in \cite{Tap_smec, Tap_chol}, the dynamics of the displacement field of cholesterics and smectics are indeed equivalent, with the layered state having long-range order in three dimensions for $\zeta<0$ (unlike their equilibrium counterparts which only have quasi-long-range order in three dimensions) and being unstable for $\zeta>0$. However, the hydrodynamics of active cholesterics differs crucially from that of smectics through the effect of the chiral active force on the velocity field \SR{\emph{in the plane of}} the layers. 

The vortical flow caused by the chiral active force can be used to control and create a vortex lattice state \cite{Vitelli} with a well defined lattice constant in an active cholesteric system. This hinges on a mapping between an \emph{externally} imposed stress and an internal and active \emph{achiral} stress. An external stress can be imposed via a free energy term $F_{ext}[u]=\int\sigma_0 E$, which gives rise to a force $-\sigma_0\nabla\cdot(\boldsymbol{w}+E{\bsf I})$~\cite{supp}. As a consequence an achiral active stress acts identically to an \emph{external} stress with $\sigma_0=\zeta$ up to an isotropic piece which can be absorbed into the pressure in an incompressible system and the instability of an \emph{active} layered state for $\zeta>0$ maps onto the Helfrich-Hurault instability of a passive layered state under dilative stress~\cite{clark1973,delaye1973}. In the externally-stressed instability a square lattice undulated pattern $u=u_0\cos q_px\cos q_py$ (an egg-crate-like structure) is realised~\cite{delrieu1974,senyuk2006} and because of this mapping, the same pattern should be realised beyond the achiral active instability as well. Due to the chiral active force $\propto z_c$, the egg-crate-like undulation leads to an in-plane vorticity $\Omega_z \propto (z_c q_p^2 u_0 / \eta) \cos q_p x \cos q_p y$, arising spontaneously from the active instability. This is the vortex lattice depicted in Fig.~\ref{fig:3dHH}. 

The correspondence between an external stress and the active achiral stress allows for a quantitative measurement of the activity. The critical threshold for a layered state of finite extent $d$ in the $z$ direction is $\zeta + \sigma_0 = (2\pi/d)\sqrt{BK}$, with the instability setting in at wavevector $q_p \approx (\pi^2 B/4d^2 K)^{1/4}$. When $\zeta$ is negative the layered state is stable to the activity and a dilatative external stress, $\sigma_0 > 0$, can be applied till the \SR{Helfrich-Hurault instability sets in \cite{commHH}. This measures the active stress: $\zeta=(2\pi/d)\sqrt{BK}-\sigma^{cr}_0$.} Conversely, for $\zeta>0$ the smallest $|\sigma_0|$ that suppresses the spontaneous Helfrich-Hurault instability, $\sigma_0^{cr}$ yields the active stress: $\zeta=|\sigma_0^{cr}|+(2\pi/d)\sqrt{BK}$ from which we can calculate the achiral active stress strength from the knowledge of the bulk and the bending modulus. The magnitude of the vorticity is $\propto z_c/\eta$. The chiral active force should scale as $z_c\sim \zeta\ell$ where $\ell$ is the length of an elementary active unit (both $\zeta$ and $z_c$ are also likely to be functions of the concentration of the active units). Therefore, in principle, we can estimate both the chiral and achiral active stress if an active cholesteric is prepared using living liquid crystals \cite{LLC}, which can be engineered, for instance, by suffusing a \emph{passive} cholesteric with bacteria, and \AMF{obtain a vortex lattice state with a lattice constant determined by the physics governing passive Helfrich-Hurault instability in an external field \cite{senyuk2006, clark1973}.}

\AMF{We now turn to a two-dimensional layered state -- an array of lines in the $x-z$ plane with normals on average along $\hat{{\bf z}}$ -- in a chiral, internally driven fluid. As in the three-dimensional cholesteric state, we obtain the coupled dynamics of the displacement and velocity field equations to leading order in gradients:} $\partial_tu=v_z$ and
\begin{equation}
\label{vel2du}
{\eta}{{\nabla}}^2{{\bf v}}=\hat{\bf z}\frac{\delta{F}[{u}]}{\delta{u}}+{\nabla}{ p}+{\nabla}\cdot
({\zeta} \boldsymbol{w} - {\zeta}_{c}\boldsymbol{\varepsilon}\cdot \boldsymbol{w})
\end{equation}
where, to linear order in ${u}$, $w_{zz}=-w_{xx}=\partial_zu$ and $w_{zx}= w_{xz}=\partial_xu$, $\zeta=(\psi_1^{0}q_s)^2\zeta_H$ and $\zeta_c=-(\psi_1^{0}q_s)^2\bar{\zeta}_c$ \cite{supp}. 
In \eqref{vel2du}, \AMF{the chiral active force $\propto\zeta_c$ appears at the \emph{same} order in gradients as the achiral active force, unlike in three-dimensional cholesterics. Further, again unlike in three-dimensional cholesterics, it will be shown to affect the displacement field dynamics at linear order. In fact, the term $\propto\zeta_c$ is fundamentally distinct from the force $\propto z_c$ in \eqref{vel3du}; it is not obtained by averaging a thin $x-z$ slice of a three-dimensional cholesteric. It leads to a chiral active force \textit{along} the layers in response to both curvature and compression of the layers i.e., a pure $\hat{{\bf z}}$ deformation leads to a force along $\hat{\bf y}$, in a direction determined by the sign of $\zeta_c$, which is only possible since the film has a distinguished normal, breaking three-dimensional rotation invariance, and the layered state breaks rotation (and translation) invariance in the plane of the film.} This effect is related to the odd elasticity \cite{Deboo2} of chiral active solids, but is odder still. A smectic breaks translation invariance only along one direction, so the (linearised) strain is simply $\bigl[ \begin{smallmatrix} 0 & 0 \\ 0 & \partial_z u \end{smallmatrix} \bigr]$. Ordinary odd elasticity would create a stress $\propto \bigl[ \begin{smallmatrix} 0 & \partial_z u \\ 0 & 0 \end{smallmatrix} \bigr]$ acting along the layers of the smectic in a direction where there is no elastic mode. Instead, the chiral activity $\zeta_c$ produces both a \textit{simple} shear stress $\sigma_{xz} = \sigma_{zx} = -\zeta_c \partial_z u$ in response to strain and also a \textit{pure} shear stress $\sigma_{xx} = -\sigma_{zz} =- \zeta_c \partial_x u$ in response to tilt. The chiral active stress implies that a localised compression of the layer spacing produces a shear flow parallel to the layers. 

We now demonstrate that a periodic array of lines in a two-dimensional film is generically destabilised due to the chiral active force. Eliminating the pressure using the incompressibility constraint in \eqref{vel2du}, solving for the velocity field and writing the wavevector ${\bf q}\equiv(q_x,q_z)=q(\sin\theta_q,\cos\theta_q)$ where $\theta_q$ is the angle between the layer normal and the wavevector, we obtain the eigenfrequency to $\mathcal{O}(q^0)$ 
\begin{equation}
\label{linstab2d}
\omega=-\frac{i}{4{\eta}}\left({{B}}\sin^2 2\theta_q-4{\zeta}\sin^2 \theta_q-2{{\zeta}_c}\sin 2\theta_q\right) + O(q^2) . 
\end{equation}
This implies an instability of the layered state for wavevector direction $\theta_q$ just above (just below) zero for ${\zeta}_c>0$ ($<0$). This generic chiral instability for either sign of $\zeta_c$ is distinct from the spontaneous Helfrich-Hurault \cite{deGen} instability of active smectics or cholesterics, which is achiral, arises for positive $\zeta$ \cite{Tap_smec, Tap_chol}, and grows fastest at $\theta_q\approx \pi/2$. 
Eq. \eqref{linstab2d} implies that in a system confined at a scale $d$ along $\hat{{\bf x}}$ so that the smallest $q_x \sim 1/d$, the minimum value of the chiral active stress for which the layered state is unstable $\sim 1/d$. Further, this instability requires both momentum conservation \emph{and} incompressibility. It is eliminated if the film is supported on a substrate which would add a wavevector-independent damping $-\Gamma {\bf v}$ to \eqref{vel2du}. The eigenfrequency for the displacement fluctuations then vanishes at small $q$ as $\mathcal{O}(q^2)$, and permeative \cite{deGen} terms $\nabla \nabla u$ in the displacement equation, subdominant for a free-standing film, now enter at the same order in gradients. Of these, terms $\propto \partial_z^2u$ are crucial, while others can be absorbed into redefinitions of $\zeta$ and $\zeta_c$). The resulting eigenfrequency is 
\begin{equation}
\label{linstabsubstrate}
\omega= -\frac{iq^2}{4\Gamma} \left({{B}}\sin^2 2\theta_q+\mathcal{B}\cos^2\theta_q-4{\zeta}\sin^2 \theta_q-2{{\zeta}_c}\sin 2\theta_q\right)
\end{equation}
where $\mathcal{B}$ is the coefficient of the $\partial_z^2u$ permeative term multiplied by the friction coefficient, and the instability now occurs only if $|\zeta_c|>(B/2)+\mathcal{B}/2-\zeta$, for directions $\theta_q\approx\pi/4$. Compressibility, as in a film bounded by bulk fluid\AMNNN{, at large enough scales} \cite{Cai}, is also stabilizing \cite{Ano_LRO}. A detailed solution \cite{supp} of the tangent-plane velocity in this case leads to the eigenfrequency for the displacement field
\begin{equation}
\label{linstabcomp}
\omega=-\frac{i|{q}|}{4{\eta}}\left[({B}\cos^2\theta_q-\zeta)(1+\sin^2\theta_q)-{1}{2}{\zeta}_c\sin2\theta_q\right]
\end{equation}
Equation \eqref{linstabcomp} yields an instability if $|\zeta_c|>[(B/2)-\zeta]/8$, for $\theta_q\approx\pi/4$ irrespective of the sign of $\zeta_c$. We expand on this in \cite{supp}. \AMF{While a free-standing film with a generically unstable chiral layered state may be difficult to access experimentally, films supported on a substrate or immersed in a bulk fluid can be engineered and the instability as in \eqref{linstabsubstrate} and \eqref{linstabcomp} may be observed.}

In this paper we have developed the hydrodynamic theory of active chiral, layered states in two and three dimensions and demonstrated that the combination of internal drive and broken-parity qualitatively modifies the dynamics and stability of these phases unlike in their equilibrium counterparts. We conclude with a brief discussion of proposals for experimental realisations, applications and possible extensions. A three-dimensional active or living cholesteric can be constructed by releasing swimming bacteria into passive biocompatible cholesteric liquid crystals, yielding a system which should display the vortex-lattice state we predict. Similarly, introducing \emph{passive} chiral particles in an active but achiral fluid also leads to the chiral active forces discussed in our work, allowing for the realisation of a wide range of artificial active cholesteric materials. Furthermore, multiple biological systems display cholesteric organisation, the most spectacular of which is DNA in chromatin \cite{Francoise}, which \textit{in vivo} may be affected by DNA polymerases leading to chiral active forces of the form that we describe here. 

In addition to free surfaces or interfaces of three-dimensional materials, there are numerous possibilities for realising a two-dimensional cholesteric phase. For instance, via the melting of a chiral version \cite{Deboo2} of anisotropic active solids \cite{Ano_sol} along one direction in analogy with the emergence of (achiral) smectic phases due to an anisotropic dislocation-mediated melting of two-dimensional crystals in which dislocations unbind along one direction \cite{HalperinNelson,NelsonHalperin,Young,Ostlund,OTZ, Radz_smec_gauge} (see \cite{Vincenzo_disc} for a description of dislocations in chiral active solids). Two-dimensional layered states are also observed in active nematic fluids both in experiments on motor-microtubule gels~\cite{Dogic,Ano_und, Ramin_und} and simulations \cite{Sagues} and since these gels are known to be chiral \cite{Sano_chi}, the physics we describe for two-dimensional, chiral layered states may be observable there. Chirality has been shown to be important in epithelial cell layers \cite{Carles} and a density modulated phase in these systems will lead to another realisation of two dimensional cholesterics. Non-mutual, two-species Cahn-Hilliard models \cite{Suronr, Cristinanr} also spontaneously form banded phases and chiral variants of these models would lead to two-dimensional chiral layered states. 
Finally, two-dimensional smectic phases in parity-broken systems are also possible in two-dimensional electron gases~\cite{QHS} and, when they are irradiated by microwave radiation \cite{Alicea}, may have a dynamics equivalent to the one described here. 
Therefore, there are abundant possibilities for creating two- and three-dimensional active cholesteric states. 

S.R. acknowledges research support from a J.C. Bose Fellowship of the SERB, India and from the Tata Education \& Development Trust. We  acknowledge illuminating discussions during the KITP 2020 online program on Symmetry, Thermodynamics and Topology in Active Matter. This research was supported in part by the National Science Foundation under Grant No. NSF PHY-1748958.


\begin{thebibliography}{}
\bibitem{TGB} S.R. Renn and T.C. Lubensky, Abrikosov dislocation lattice in a model of the cholesteric--to--smectic-A transition, Phys. Rev. A {\bf 38}, 2132 (1988). \doi{10.1103/PhysRevA.38.2132}
\bibitem{Mermin_RMP} D.C. Wright and N.D. Mermin, Crystalline liquids: the blue phases, Rev. Mod. Phys. {\bf 61}, 385 (1989). \doi{10.1103/RevModPhys.61.385}
\bibitem{deGen} P.G. de Gennes and J. Prost, The Physics of Liquid Crystals (second edition), Clarendon, Oxford (1993). 
\bibitem{opticsfoot} Chirality is of course manifest in the bulk optical and electrical properties of cholesterics.
\bibitem{Lub_chol} T.C. Lubensky, Phys. Rev. A {\bf 6}, 452 (1972). \doi{10.1103/PhysRevA.6.452}
\bibitem{Rad_Lub} L. Radzihovsky and  T.C. Lubensky, Nonlinear smectic elasticity of helical state in cholesteric liquid crystals and helimagnets, Phys. Rev. E {\bf 83}, 051701 (2011). \doi{10.1103/PhysRevE.83.051701}
\bibitem{Tap_chol} C.A. Whitfield, T.C. Adhyapak, A. Tiribocchi, G.P. Alexander, D. Marenduzzo, and S. Ramaswamy, Hydrodynamic instabilities in active cholesteric liquid crystals, Eur. Phys. J. E {\bf 40}, 50 (2017). \doi{10.1140/epje/i2017-11536-2} 
\bibitem{Tap_smec} T.C. Adhyapak, S. Ramaswamy, and  J. Toner, Live Soap: Stability, Order, and Fluctuations in Apolar Active Smectics, Phys. Rev. Lett. {\bf 110}, 118102 (2013). \doi{10.1103/PhysRevLett.110.118102}
\bibitem{Deboo2} C. Scheibner, A. Souslov, D. Banerjee, P. Sur\'owka, W.T.M. Irvine, and V. Vitelli, Odd elasticity. Nat. Phys. {\bf 16}, 475 (2020). \doi{10.1038/s41567-020-0795-y}
\bibitem{SRJSTAT} S. Ramaswamy, Active Matter, J. Stat. Mech. (2017) 054002. \doi{10.1088/1742-5468/aa6bc5}
\bibitem{LPDJSTAT} L.P. Dadhichi, A. Maitra, and S. Ramaswamy,  Origins and diagnostics of the nonequilibrium character of active systems, J. Stat. Mech. (2018) 123201. \doi{10.1088/1742-5468/aae852}
\bibitem{RMP} M.C. Marchetti, J-F. Joanny, S. Ramaswamy, T.B. Liverpool, J. Prost, M. Rao, and R.A. Simha, Rev. Mod. Phys. {\bf 85}, 1143 (2013). \doi{10.1103/RevModPhys.85.1143}
\bibitem{SRrev} S. Ramaswamy, The Mechanics and Statistics of Active Matter, Annu. Rev. Condens. Matter Phys. {\bf 1}, 323 (2010). \doi{10.1146/annurev-conmatphys-070909-104101}
\bibitem{Prost_nat} J. Prost, F. J\"ulicher, and J-F. Joanny, Active gel physics. Nat. Phys. {\bf 11}, 111 (2015). \doi{10.1038/nphys3224}
\bibitem{SalJul} G. Salbreux and F. J\"ulicher, Mechanics of active surfaces, Phys. Rev. E {\bf 96}, 032404 (2017). \doi{10.1103/PhysRevE.96.032404}
\bibitem{TonTuRam} J. Toner, Y. Tu, and S. Ramaswamy, Hydrodynamics and phases of flocks, Ann. Phys. {\bf 318}, 170 (2005). \doi{10.1016/j.aop.2005.04.011}
\bibitem{CatesRev} M.E. Cates and J. Tailleur, Motility-Induced Phase Separation, Annu. Rev. Condens. Matter Phys. {\bf 6}, 219 (2015). \doi{10.1146/annurev-conmatphys-031214-014710}
\bibitem{Curierep} K. Kruse, J-F. Joanny, F. J\"ulicher, J. Prost, and K. Sekimoto, Generic theory of active polar gels: a paradigm for cytoskeletal dynamics, Eur. Phys. J. E {\bf 16}, 5 (2005). \doi{10.1140/epje/e2005-00002-5}
\bibitem{Shelleyrev} D. Saintillan and M.J. Shelley, Active suspensions and their nonlinear models, Comptes Rendus Physique {\bf 14}, 497 (2013). \doi{10.1016/j.crhy.2013.04.001}
\bibitem{scalar} A.P. Solon, J. Stenhammar, M.E. Cates, Y. Kafri, and J. Tailleur, Generalized thermodynamics of phase equilibria in scalar active matter, Phys. Rev. E {\bf 97}, 020602 (2018). \doi{10.1103/PhysRevE.97.020602}
\bibitem{smecp} P.  Romanczuk, H. Chat\'e, L. Chen, S. Ngo, and J. Toner, Emergent smectic order in simple active particle models, New J. Phys. {\bf 18}, 063015 (2016). \doi{10.1088/1367-2630/18/6/063015}
\bibitem{Chen_Toner} L. Chen and J. Toner, Universality for Moving Stripes: A Hydrodynamic Theory of Polar Active Smectics, Phys. Rev. Lett. {\bf 111}, 088701 (2013). \doi{10.1103/PhysRevLett.111.088701}
\bibitem{Ano_sol} A. Maitra and S. Ramaswamy, Oriented Active Solids, Phys. Rev. Lett. {\bf 123}, 238001 (2019). \doi{10.1103/PhysRevLett.123.238001} 
\bibitem{Lewis1} B. Liebchen and D. Levis, Collective Behavior of Chiral Active Matter: Pattern Formation and Enhanced Flocking, Phys. Rev. Lett. {\bf 119}, 058002 (2017). \doi{10.1103/PhysRevLett.119.058002}
\bibitem{Lewis2}  D. Levis and B. Liebchen, Micro-flock patterns and macro-clusters in chiral active Brownian disks, J. Phys.: Condens. Matter {\bf 30}, 084001 (2018). \doi{10.1088/1361-648x/aaa5ec}
\bibitem{Strempel} S. F\"urthauer, M. Strempel, S.W. Grill, F. J\"ulicher, Active chiral fluids, Eur. Phys. J. E {\bf 35}, 89 (2012). \doi{10.1140/epje/i2012-12089-6}
\bibitem{Cates_drop} E. Tjhung, M.E. Cates, and D. Marenduzzo, Contractile and chiral activities codetermine the helicity of swimming droplet trajectories, Proc. Natl. Acad. Sci. U.S.A. {\bf 114}, 4631 (2017). \doi{10.1073/pnas.1619960114 } 
\bibitem{Ano_chi} A. Maitra, M. Lenz, and R. Voituriez, Chiral Active Hexatics: Giant Number Fluctuations, Waves, and Destruction of Order, Phys. Rev. Lett. {\bf 125}, 238005 (2020). \doi{10.1103/PhysRevLett.125.238005}
\bibitem{Luca_chi}  L.A. Hoffmann, K. Schakenraad, R.M.H. Merks, and L. Giomi, Soft Matter {\bf 16}, 764 (2020). \doi{10.1039/C9SM01851D}
\bibitem{Carles} G. Duclos, C. Blanch-Mercader, V. Yashunsky, G. Salbreux, J-F. Joanny, J. Prost, and P. Silberzan, Spontaneous shear flow in confined cellular nematics, Nat. Phys. {\bf 14}, 728 (2018). \doi{10.1038/s41567-018-0099-7}
\bibitem{Cates_chi} T. Markovich, E. Tjhung, and M.E. Cates, Chiral active matter: microscopic `torque dipoles' have more than one hydrodynamic description, New J. Phys. {\bf 21}, 112001 (2019). \doi{10.1088/1367-2630/ab54af}
\bibitem{Seb1} S. F\"urthauer, M. Strempel, S.W. Grill, and F. J\"ulicher, Active Chiral Processes in Thin Films, Phys. Rev. Lett. {\bf 110}, 048103 (2013). \doi{10.1103/PhysRevLett.110.048103}
\bibitem{HohenCross} M.C. Cross and P.C. Hohenberg, Pattern formation outside of equilibrium, Rev. Mod. Phys. {\bf 65}, 851 (1993). \doi{10.1103/RevModPhys.65.851}
\bibitem{CatesH1} A. Tiribocchi, R. Wittkowski, D. Marenduzzo, and M.E. Cates, Active Model H: Scalar Active Matter in a Momentum-Conserving Fluid, Phys. Rev. Lett. {\bf 115}, 188302 (2015). \doi{10.1103/PhysRevLett.115.188302} 
\bibitem{CatesH2} R. Singh and  M.E. Cates, Hydrodynamically Interrupted Droplet Growth in Scalar Active Matter, Phys. Rev. Lett. {\bf 123}, 148005 (2019). \doi{10.1103/PhysRevLett.123.148005}
\bibitem{deGroot}  S.R. De Groot and P. Mazur, Non-Equilibrium Thermodynamics, Dover Books on Physics (2011). 
\bibitem{AMB1} R. Wittkowski, A. Tiribocchi, J. Stenhammar, R.J. Allen, D. Marenduzzo, and M.E. Cates, Scalar $\phi^{4}$ Field Theory for Active-Particle Phase Separation, Nat. Commun. {\bf 5}, 4351 (2014). \doi{10.1038/ncomms5351}
\bibitem{AMB2} A.P. Solon, M.E. Cates, and J. Tailleur, Active brownian particles and run-and-tumble particles: A comparative study, Eur. Phys. J. Spec. Top. {\bf 224}, 1231 (2015). \doi{10.1140/epjst/e2015-02457-0}
\bibitem{Cesare1} C. Nardini, \'E. Fodor, E. Tjhung, F. v. Wijland, J. Tailleur, and M.E. Cates, Entropy Production in Field Theories without Time-Reversal Symmetry: Quantifying the Non-Equilibrium Character of Active Matter, Phys. Rev. X {\bf 7}, 021007 (2017). \doi{10.1103/PhysRevX.7.021007}
\bibitem{CesarePRX} E. Tjhung, C. Nardini, and M.E. Cates, Cluster Phases and Bubbly Phase Separation in Active Fluids: Reversal of the Ostwald Process, Phys. Rev. X {\bf 8}, 031080 (2018). \doi{10.1103/PhysRevX.8.031080}
\bibitem{supp} Supplementary material. 
\bibitem{MPP} P.C. Martin, O. Parodi, and P.S. Pershan, Unified Hydrodynamic Theory for Crystals, Liquid Crystals, and Normal Fluids, Phys. Rev. A {\bf 6}, 2401 (1972). \doi{10.1103/PhysRevA.6.2401} 
\bibitem{StarkLubensky} H. Stark and T.C. Lubensky, Poisson bracket approach to the dynamics of nematic liquid crystals: The role of spin angular momentum, Phys. Rev. E {\bf 72}, 051714 (2005). \doi{10.1103/PhysRevE.72.051714} 
\bibitem{SH} J. Swift and P.C. Hohenberg, Hydrodynamic fluctuations at the convective instability, Phys. Rev. A {\bf 15}, 319 (1977). \doi{10.1103/PhysRevA.15.319}
\bibitem{GrinPel} G. Grinstein and  R.A. Pelcovits, Nonlinear elastic theory of smectic liquid crystals, Phys. Rev. A {\bf 26}, 915 (1982). \doi{10.1103/PhysRevA.26.915}
\bibitem{Vitelli} A. Souslov, B. van Zuiden, D. Bartolo, and V. Vitelli,  Topological sound in active-liquid metamaterials. Nat. Phys. {\bf 13}, 1091 (2017).\doi{/10.1038/nphys4193}
\bibitem{clark1973} N.A. Clark and R.B. Meyer, Strain-induced instability of monodomain smectic A and cholesteric liquid crystals, Appl. Phys. Lett. {\bf 22}, 493 (1973). \doi{10.1063/1.1654481} 
\bibitem{delaye1973} M. Delaye, R. Ribotta, and G. Durand, Buckling instability of the layers in a smectic-A liquid crystal, Phys. Lett. A {\bf 44}, 139 (1973). \doi{10.1016/0375-9601(73)90822-0} 
\bibitem{delrieu1974} J.M. Delrieu, Comparison between square, triangular, or one dimensional lattice of distortions in smectic A and cholesteric crystals for superposed strain and magnetic field of any directions, J. Chem. Phys. {\bf 60}, 1081 (1974). \doi{10.1063/1.1681116} 
\bibitem{senyuk2006} B.I. Senyuk, I.I. Smalyukh, and O.D. Lavrentovich, Undulations of lamellar liquid crystals in cells with finite surface anchoring near and well above the threshold, Phys. Rev. E {\bf 74}, 011712 (2006). \doi{10.1103/PhysRevE.74.011712} 
\bibitem{commHH} \AMF{In practice, this external dilative stress is often imposed by the means of an external electric or magnetic field.}
\bibitem{LLC} S. Zhou, A. Sokolov, O.D. Lavrentovich, and I.S. Aranson, Living liquid crystals, Proc. Natl. Acad. Sci. U.S.A. {\bf 111}, 1265 (2014). \doi{10.1073/pnas.1321926111} 
\bibitem{Cai} W. Cai and T.C. Lubensky, Hydrodynamics and dynamic fluctuations of fluid membranes, Phys. Rev. E {\bf 52}, 4251 (1995). \doi{10.1103/PhysRevE.52.4251} 
\bibitem{Ano_LRO} A. Maitra, Long-range uniaxial order in active fluids at two-dimensional interfaces, arXiv:2010.15044.
\bibitem{Francoise} A. Leforestier and F. Livolant, Cholesteric liquid crystalline DNA; a comparative analysis of cryofixation methods, Biology of the Cell {\bf 71} 115 (1991).
\bibitem{HalperinNelson} B.I. Halperin and D.R. Nelson, Theory of Two-Dimensional Melting, Phys. Rev. Lett. {\bf 41}, 121 (1978). \doi{10.1103/PhysRevLett.41.121} 
\bibitem{NelsonHalperin} D.R. Nelson and B.I. Halperin, Dislocation-mediated melting in two dimensions, Phys. Rev. B {\bf 19}, 2457 (1979). \doi{10.1103/PhysRevB.19.2457}
\bibitem{Young} A.P. Young, Melting and the vector Coulomb gas in two dimensions, Phys. Rev. B {\bf 19}, 1855 (1979). \doi{10.1103/PhysRevB.19.1855} 
\bibitem{Ostlund} S. Ostlund and B.I. Halperin, Dislocation-mediated melting of anisotropic layers, Phys. Rev. B {\bf 23}, 335 (1981). \doi{10.1103/PhysRevB.23.335}
\bibitem{OTZ} S. Ostlund, J. Toner, and A. Zippelius, Dynamics of 2D anisotropic solids, smectics and nematics, Ann. Phys. {\bf 144}, 345 (1982). \doi{10.1016/0003-4916(82)90118-X}
\bibitem{Radz_smec_gauge}  L. Radzihovsky, Quantum smectic gauge theory, arXiv:2009.06632. 
\bibitem{Vincenzo_disc} L. Braverman, C. Scheibner, and V. Vitelli, Topological defects in non-reciprocal active solids with odd elasticity, arXiv:2011.11543.
\bibitem{Dogic} T. Sanchez, D. Chen, S.J. DeCamp, M. Heymann, and Z. Dogic,  Spontaneous motion in hierarchically assembled active matter, Nature {\bf 491}, 431 (2012). \doi{10.1038/nature11591}
\bibitem{Ano_und} A. Senoussi, S. Kashida, R. Voituriez, J.C. Galas, A. Maitra, and A. Estevez-Torres, Tunable corrugated patterns in an active nematic sheet, Proc. Natl. Acad. Sci. U.S.A. {\bf 116}, 22464 (2019). \doi{10.1073/pnas.1912223116}
\bibitem{Ramin_und} T. Str\"ubing, A. Khosravanizadeh, A. Vilfan, E. Bodenschatz, R. Golestanian, and I. Guido, Wrinkling Instability in 3D Active Nematics, Nano Lett. {\bf 20}, 6281 (2020). \doi{10.1021/acs.nanolett.0c01546}
\bibitem{Sagues} A. Doostmohammadi, J. Ignes-Mullol, J.M. Yeomans, and  F. Sagues, Active nematics, Nat. Commun. {\bf 9}, 3246 (2018). \doi{10.1038/s41467-018-05666-8}
\bibitem{Sano_chi} T. Yamamoto and M. Sano, Theoretical model of chirality-induced helical self-propulsion, Phys. Rev. E {\bf 97}, 012607 (2018). \doi{10.1103/PhysRevE.97.012607}
\bibitem{Suronr} S. Saha, J. Agudo-Canalejo, and R. Golestanian, Scalar Active Mixtures: The Nonreciprocal Cahn-Hilliard Model, Phys. Rev. X {\bf 10}, 041009 (2020). \doi{10.1103/PhysRevX.10.041009}
\bibitem{Cristinanr} Z. You, A. Baskaran, and M.C. Marchetti, Nonreciprocity as a generic route to traveling states, Proc. Natl. Acad. Sci. U.S.A. {\bf 117}, 19767 (2020). \doi{10.1073/pnas.2010318117}
\bibitem{QHS} M.J. Lawler and E. Fradkin, Quantum Hall smectics, sliding symmetry, and the renormalization group, Phys. Rev. B {\bf 70}, 165310 (2004). \doi{10.1103/PhysRevB.70.165310}
\bibitem{Alicea} J. Alicea, L. Balents and M.P.A. Fisher, A. Paramekanti, and L. Radzihovsky, Transition to zero resistance in a two-dimensional electron gas driven with microwaves, Phys. Rev. B {\bf 71}, 235322 (2005). \doi{10.1103/PhysRevB.71.235322}
\end{thebibliography}
\end{document}



\title{Active cholesterics: odder than odd elasticity: Supplementary Material}

\author{S. J. Kole}
\email{swapnilkole@iisc.ac.in}
\affiliation{Centre for Condensed Matter Theory, Department of Physics, Indian Institute of Science, Bangalore 560 012, India}
\author{Gareth P. Alexander}
\email{G.P.Alexander@warwick.ac.uk}
\affiliation{Department of Physics and Centre for Complexity Science, University of Warwick, Coventry CV4 7AL, United Kingdom}
\author{Sriram Ramaswamy}
\email{sriram@iisc.ac.in}
\affiliation{Centre for Condensed Matter Theory, Department of Physics, Indian Institute of Science, Bangalore 560 012, India}
\author{Ananyo Maitra}
\email{nyomaitra07@gmail.com}
\affiliation{Sorbonne Universit\'{e} and CNRS, Laboratoire Jean Perrin, F-75005, Paris, France}


\begin{abstract}
In this supplement we present the detailed calculations corresponding to the results in the main text. In Sec. \ref{sec1} we discuss active theories of conserved and non-conserved chiral, scalar order parameters in momentum-conserved system. We discuss the conserved order parameter case in Sec. \ref{sec1sub1}, first in two dimensions in Sec. \ref{2dH} and then in three in Sec. \ref{3dH}. Then, in Sec. \ref{sec1sub2}, we discuss the theory of a chiral, scalar non-conserved order parameter in a momentum conserved fluid, again, first in two dimensions in Sec. \ref{2dA} and then in three in Sec. \ref{3dA}. 
In Sec. \ref{sec2} we use a free-energy of the Swift-Hohenberg form to obtain the hydrodynamic equations for layered states in two and three dimensions. We first consider it in a two-dimensional system in Sec. \ref{2dlayer} and obtain the hydrodynamic equations first starting from the conserved model in Sec. \ref{2dconslayer} and then from the non-conserved model in Sec. \ref{2dlaynoncons}. We then consider a three-dimensional system in Sec. \ref{3dlayer} again first starting from the conserved order parameter in Sec. \ref{3dconslayer} and then from the non-conserved order parameter model in Sec. \ref{3dlaynoncons}. Then in Sec. \ref{Eqastrestr} we demonstrate that the effect of the achiral active stress in a layered state can be compensated by an external stress. In Sec. \ref{equiallstr} we show that various distinct ways of introducing activity in a layered system are equivalent to the theory we construct. In Sec. \ref{linchisec} we consider the linear theory of chiral, layered states in two dimensions in Sec. \ref{linchisec2d} ans in three dimensions in Sec. \ref{linchisec3d}. Finally, in Sec. \ref{beyondlin} we demonstrate that a three-dimensional chiral layered state can be used to controllably generate vortex lattice phases.
\end{abstract}


\keywords{Chirality, activity}
\maketitle

\section{Active chiral, scalar fields in momentum conserved systems}
\label{sec1}
In this section, we construct the generic hydrodynamic theories of a conserved and non-conserved scalar order parameter field in a chiral, momentum-conserved system. We start with the conserved case and then proceed to the non-conserved one. 
\subsection{Conserved order parameter: Active model H*}
\label{sec1sub1}
For a conserved order parameter, our theory is an extension of the active model H (itself an extension of model H in the classification of \cite{HalpHohen} to active systems) \cite{CatesH1, CatesH2} to include chirality and we dub it active model H*. Like the usual model H, this will be characterised by a conserved density or order parameter field and an incompressible velocity field. Just as model H describes binary phase separation in passive systems, active model H* may describe binary phase separation in active, homochiral systems.
We will construct two distinct variants of this model. For the first variant, we consider a two-dimensional chiral suspension with a distinguished normal direction -- i.e., a film that breaks up-down symmetry. For the second variant we consider a three-dimensional momentum-conserved system. 
\subsubsection{Two-dimensional film}
\label{2dH}
We first consider a suspended thin-film of chiral, active particles in a fluid which breaks up-down symmetry, i.e., has a distinguished normal direction ${\bf N}$ which we take to be along the $\hat{\bf y}$ direction. We further define a two-dimensional antisymmetric tensor $\varepsilon_{ij}=\epsilon_{ikj}N_k$ (i.e., $\varepsilon_{xz}=1$ and $\varepsilon_{zx}=-1$ and other components are $0$). With these definitions, the equation of motion for the two-dimensional concentration or conserved order parameter field of the chiral, active particles $\psi({\bf r}, t)$, where ${\bf r}\equiv(x,z)$, is
\begin{equation}
\label{psieq2d}
\partial_t\psi=-\nabla\cdot(\psi{\bf v})+{M}\nabla^2\frac{\delta  F [\psi]}{\delta\psi}+\nabla\cdot{{\bf J}}_a+\nabla\cdot{{\bf J}}_c+{{\xi}}_\psi,
\end{equation}
where $\nabla\equiv\partial_x\hat{\bf x}+\partial_z\hat{\bf z}$ denotes the two-dimensional gradient, ${{\bf J}}_a$ contains active achiral currents also present in active model H and active model B \cite{CatesH1, CatesH2, AMB1, AMB2, Cesare1}, ${{\bf J}}_c$ contains active and passive chiral currents, $ F [\psi]$ is a phenomenological free energy which would have governed the relaxation to the equilibrium state in the absence of activity and ${\xi}_\psi$ is a conserving, Gaussian white noise with the correlation $\langle{\xi}_\psi({\bf r}, t){\xi}_\psi({\bf r}', t')\rangle=-2{D}\nabla^2\delta({\bf r}-{\bf r}')\delta(t-t')$. The active, achiral currents are 
\begin{equation}
\label{acpsi2}
{{\bf J}}_a=\lambda_1\,\psi\nabla\psi(\nabla^2\psi)+\lambda_2\,\psi\nabla(\nabla\psi)^2.
\end{equation}
To lowest order in gradients, the chiral current is 
\begin{equation}
\label{ccpsi2}
{\bf J}_c=\omega_v\psi\nabla(\boldsymbol{\varepsilon}:\nabla{\bf v})+\omega_1\psi\nabla^2\psi\boldsymbol{\varepsilon}\cdot\nabla\psi+\omega_2\psi\boldsymbol{\varepsilon}\cdot\nabla(\nabla\psi)^2.
\end{equation}
The first term is a chiral, reactive coupling to the velocity field which can exist even in passive momentum-conserved systems. The second and third are chiral and active density currents which may be present even in active model B*. 
The equation of motion for the two-dimensional in-plane Stokesian velocity field is
\begin{equation}
\label{vel2dpsi}
\eta{\nabla}^2{\bf v}=\psi\nabla\frac{\delta  F [\psi]}{\delta \psi}+\nabla p-\SK{\nabla\cdot\big[(\zeta_H{\bsf I}+\bar{\zeta}_c\boldsymbol{\varepsilon})\cdot(\nabla\psi\nabla\psi)^{ST}\big]}+\omega_v(\boldsymbol{\varepsilon}\cdot\nabla)\nabla\cdot\left[ \psi\nabla\frac{\delta  F [\psi]}{\delta\psi}\right]+{\boldsymbol{\xi}}_v,
\end{equation}
where the superscript $ST$ denotes symmetrisation and trace-removal, ${\bsf I}$ is the two-dimensional identity tensor, $ p$ is the pressure enforcing the incompressibility constraint $\nabla\cdot{\bf v}=0$, $\zeta_H$ is the coefficient of the \emph{achiral} active force \cite{CatesH2, modH, FinlayScr}, $\bar{\zeta}_c$ is the coefficient of a \emph{chiral} active force whose form is similar to the ones in \cite{Pascal, Ano_chi, Luca_chi, Deboo2} and ${\boldsymbol{\xi}}_v$ is a conserving Gaussian white noise with the correlation $\langle{\boldsymbol{\xi}}_v({\bf r},t){\boldsymbol{\xi}}_v({\bf r}',t')\rangle=-2{\bsf I}{D}_v\nabla^2\delta({\bf r}-{\bf r}')\delta(t-t')$. In the passive limit, the noise strengths must satisfy the relation ${D}/{M}={D}_v/\eta$. The active forces with the coefficients $\zeta_H$ and $\bar{\zeta}_c$ are the only ones allowed at this order in gradients and fields. Eqs. \eqref{psieq2d} and \eqref{vel2dpsi} are the general dynamical equations for active model H* in a two-dimensional up-down symmetry-broken film. Only the form of the free energy $ F [\psi]$ needs to be specified to complete the description. There is no explicitly chiral term in the free energy up to very high orders in gradients and fields (the first explicitly chiral term contains \emph{nine} gradients). Therefore, if we wish to describe classic liquid-gas phase separation, a standard $\phi^4$ free energy is sufficient. In this paper we are concerned not with binary phase separation but one-dimensional periodic states for which we will use a standard Swift-Hohenberg free energy \cite{SH, HohenCross, Brazovskii}.

\subsubsection{Three-dimensional systems}
\label{3dH}
We now consider a bulk, three-dimensional system with three-dimensional density $\psi({\bf r}, t)$ and velocity ${\bf v}({\bf r}, t)$ fields where ${\bf r}=(x,y,z)$. The dynamical equation for $\psi$
\begin{equation}
\label{psieq3d}
\partial_t\psi=-\nabla\cdot(\SK{\psi}{\bf v})+M\nabla^2\frac{\delta F[\psi]}{\delta\psi}+\nabla\cdot{\bf J}_a+\nabla\cdot{\bf J}_c+{\xi}_\psi,
\end{equation}
with $\langle\xi_\psi({\bf r}, t)\xi_\psi({\bf r}', t')\rangle=-2D\nabla^2\delta({\bf r}-{\bf r}')\delta(t-t')$, $\nabla$ being the three-dimensional gradient operator and 
\begin{equation}
{\bf J}_a=\lambda_1\psi\nabla\psi\nabla^2\psi+\lambda_2\psi\nabla(\nabla\psi)^2
\end{equation}
having the same form as the two-dimensional model but the chiral current is fundamentally modified:
\begin{equation}
{\bf J}_c=\Omega_v\psi\nabla^2(\nabla\times{\bf v}).
\end{equation}
This is a chiral coupling to the velocity also allowed in passive chiral systems and was discussed in \cite{Andreev} (the other chiral current discussed in \cite{Andreev} is nonlinear in the velocity field and we don't consider it in this paper). Note that unlike in the two-dimensional films, here we have not introduced a chiral density current. The first such current appears at 
sixth order in gradients. Finally, the constitutive equation for the Stokesian velocity field is
\begin{equation}
\label{vel3dpsi}
\eta\nabla^2{v}_i=\psi\partial_i\frac{\delta F[\psi]}{\delta \psi}+\partial_i p-\partial_j[{\zeta}_H\partial_i\psi\partial_j\psi+ \bar{z}_c \partial_l(\epsilon_{ijk}\partial_k\psi\partial_l\psi)]-\Omega_v\epsilon_{ijk}\partial_j\partial_l\partial_l\left[\psi\partial_k\frac{\delta F[\psi]}{\delta\psi}\right]+\xi_{v_i},
\end{equation}
where $\langle{{\xi}}_{v_i}({\bf r},t){{\xi}}_{v_j}({\bf r}',t')\rangle=-2\delta_{ij}{D}_v{\nabla}^2\delta({\bf r}-{\bf r}')\delta(t-t')$ and $ p$ is the pressure that enforces the three-dimensional incompressibility constraint $\nabla\cdot{\bf v}=0$. In equilibrium, in addition to all active terms being $0$, ${D}/{M}={D}_v/{\eta}$. Unlike in a two-dimensional film, the chiral active force, with the coefficient $ \bar{z}_c $, appears at a \emph{higher} order in gradients than the achiral active force. This force has a form similar to the one used in theories of three-dimensional chiral nematic \cite{Seb_1, Seb_2, Tjhung_1, Cates_Marko} and the velocity field resulting from it can be shown to be divergence-free. Though this stress superficially seems to be antisymmetric, it is allowed in angular momentum-conserved systems; in fact, an equivalent (up to a Belinfante-Rosenfeld tensor) explicitly \emph{symmetric} stress can be constructed \cite{MPP, Cates_Marko} which yields the same velocity field:
\begin{equation}
 \bar{z}_c \partial_l(\epsilon_{ijk}\partial_k\psi\partial_l\psi)\equiv \bar{z}_c [\epsilon_{ilk}\partial_l(\partial_k\psi\partial_j\psi)+\epsilon_{jlk}\partial_l(\partial_k\psi\partial_i\psi)].
\end{equation}
This completes the dynamics of three-dimensional active model H*. As in the two-dimensional case, the free energy $F[\psi]$ may have a simple $\phi^4$ form if we use these equations to describe liquid-gas phase separation of chiral mesogens. In this paper, however, we concentrate on layered states.

\subsection{Non-conserved order parameter}
\label{sec1sub2}
We now construct the dynamics of a \emph{non-conserved} scalar order parameter in chiral, momentum conserved systems. As in the conserved case, we first discuss a two-dimensional film with broken up-down symmetry and then a bulk, three-dimensional system and again use a Swift-Hohenberg free energy.

\subsubsection{Two-dimensional film}
\label{2dA}
We consider a suspended thin-film of chiral, active particles in a fluid which breaks up-down symmetry, i.e. has a distinguished normal direction ${\bf N}\equiv\hat{\bf y}$. We define a two-dimensional antisymmetric tensor $\varepsilon_{ij}=\epsilon_{ijk}N_k$. With these definitions, the equation of motion for the two-dimensional non-conserved order parameter field $ m ({\bf r}, t)$, where ${\bf r}\equiv(x,z)$, is
\begin{multline}
\label{eqm2d}
\partial_t m +{\bf v}\cdot\nabla m =\omega_{v1}\nabla^2(\boldsymbol{\varepsilon}:\nabla{\bf v})+\omega_{v2}\nabla m \cdot\nabla(\boldsymbol{\varepsilon}:\nabla{\bf v})+\omega_1\nabla(\nabla^2 m )\cdot\boldsymbol{\varepsilon}\cdot\nabla  m +\omega_2\nabla m \cdot\boldsymbol{\varepsilon}\cdot\nabla(\nabla  m )^2+\lambda_1(\nabla m )^2\\+\lambda_2\nabla^2 m (\nabla m )^2+\lambda_3\nabla m \cdot\nabla(\nabla m )^2+\lambda_4 m \nabla^2(\nabla m )^2+\lambda_5 m (\nabla^2 m )^2+\lambda_6 m \nabla m \cdot\nabla(\nabla^2 m )-\Gamma_m\frac{\delta  F [ m ]}{\delta m }+{\xi}_m,
\end{multline}
where $\nabla\equiv\partial_x\hat{\bf x}+\partial_z\hat{\bf z}$, ${\bf v}({\bf r},t)$ is the two-dimensional velocity field and $\xi_m({\bf r}, t)$ is a non-conserving, Gaussian white noise with the correlation $\langle{\xi}_m({\bf r}, t){\xi}_m({\bf r}', t')\rangle=2{D}_m\delta({\bf r}-{\bf r}')\delta(t-t')$. The coefficients $\omega_{v1}$, $\omega_{v2}$, $\omega_1$, $\omega_2$ and $\lambda_1$ can all be functions of $m$.
The first two terms are reactive couplings to the velocity field which are allowed in passive chiral systems as well.
The terms with the coefficients $\omega_1$, $\omega_2$ and $\lambda_i$ in \eqref{eqm2d} are active. We now display the equation of motion for the two-dimensional, in-plane Stokesian velocity field:
\begin{equation}
\label{vel2dm}
\eta{\nabla}^2{\bf v}=-(\nabla m )\frac{\delta  F [ m ]}{\delta  m }+\nabla p-\nabla\cdot[\zeta_H{\bsf I}+\bar{\zeta}_c\boldsymbol{\varepsilon}]\cdot(\nabla m \nabla m )^{ST}+(\boldsymbol{\varepsilon}\cdot\nabla)\left[\nabla^2\left(\omega_{v1}\frac{\delta F [\psi]}{\delta m }\right)-\nabla\cdot\left\{
\omega_{v2}(\nabla m )\frac{\delta F [ m ]}{\delta m }\right\}\right]+{\boldsymbol{\xi}}_{v},
\end{equation}
where the superscript $ST$ denotes symmetrisation and trace-removal, ${\bsf I}$ is the two-dimensional identity tensor, $ p$ is the pressure enforcing the incompressibility constraint $\nabla\cdot{\bf v}=0$, $\zeta_H$ is the coefficient of the \emph{achiral} active force \cite{CatesH2, modH, FinlayScr}, $\bar{\zeta}_c$ is the coefficient of a \emph{chiral} active force whose form is similar to the ones in \cite{Pascal, Ano_chi, Luca_chi, Deboo2} and ${\boldsymbol{\xi}}_{v}$ is a conserving Gaussian white noise with the correlation $\langle{\boldsymbol{\xi}}_v({\bf r},t){\boldsymbol{\xi}}_v({\bf r}',t')\rangle=-2{\bsf I}{D}_v\nabla^2\delta({\bf r}-{\bf r}')\delta(t-t')$. In the passive limit, the noise strengths must satisfy the relation ${D}_m/\Gamma_m={D}_v/\eta$. The active forces with the coefficients $\zeta_H$ and $\bar{\zeta}_c$ which are equivalent to the ones in \eqref{vel2dpsi} are the only ones allowed at this order in gradients and fields even though $ m $ is a non-conserved field. Eqs. \eqref{eqm2d} and \eqref{vel2dm} describe the dynamics of a non-conserved order parameter in a chiral, momentum conserved active system along with a definition for $ F [ m ]$. To account for phase transition between a state with $m=0$ and $m\neq 0$, one would need to use a $\phi^4$ free energy. In this paper, we study a distinct question -- the fate of a layered state given \eqref{eqm2d} and \eqref{vel2dm} and for that, we will use a free energy of the Swift-Hohenberg form \cite{SH}.

\subsubsection{Three-dimensional systems}
\label{3dA}
We now consider a bulk, three-dimensional system with a three-dimensional non-conserved order parameter $m({\bf r}, t)$ and velocity field ${\bf v}({\bf r}, t)$ where ${\bf r}\equiv(x,y,z)$. The dynamical equation for $m$ is 
\begin{multline}
\label{3dmeq}
\partial_tm+{\bf v}\cdot\nabla m=\Omega_v\nabla m\cdot\nabla^2(\nabla\times{\bf v})+\lambda_1({\nabla}{m})^2+\lambda_2{\nabla}^2{m}({\nabla}{m})^2+\lambda_3{\nabla}{m}\cdot{\nabla}({\nabla}{m})^2\\+\lambda_4{m}{\nabla}^2({\nabla}{m})^2+\lambda_5{m}({\nabla}^2{m})^2+\lambda_6{m}{\nabla}{m}\cdot{\nabla}({\nabla}^2{m})-{\Gamma}_m\frac{\delta {F}[{m}]}{\delta{m}}+{\xi}_m.
\end{multline}
Here, $\nabla$ is the three-dimensional gradient operator, ${\bf v}({\bf r}, t)$ is the three-dimensional velocity field and and $\xi_m({\bf r}, t)$ is a non-conserving, Gaussian white noise with the correlation $\langle{\xi}_m({\bf r}, t){\xi}_m({\bf r}', t')\rangle=2{D}_m\delta({\bf r}-{\bf r}')\delta(t-t')$. We have not included a chiral term involving only the gradients of $m$ which appears at higher order in gradients. {The first term of the R.H.S. of \eqref{3dmeq} is allowed even in passive systems while the terms with the coefficients $\lambda_i$ are purely active and the final term is the passive relaxation}. The equation of motion for the velocity field is 
\begin{equation}
\label{vel3dm}
{\eta}{{\nabla}}^2{{v}}_i=-({\partial_i}{m})\frac{\delta {F}[{m}]}{\delta {m}}+{\partial_i}{ p}-{\partial_j}\cdot[\zeta_H({\partial_i}{m}{\partial_j}{m})+\bar{z}_c\partial_l(\epsilon_{ijk}\partial_km\partial_lm)]+\partial_j\left[\epsilon_{ijk}\partial_l\partial_l\left(\Omega_v\partial_k m\frac{\delta F[m]}{\delta m}\right)\right]+{\boldsymbol{\xi}}_{v_i},
\end{equation}
where $\langle{{\xi}}_{v_i}({\bf r},t){{\xi}}_{v_j}({\bf r}',t')\rangle=-2\delta_{ij}{D}_v{\nabla}^2\delta({\bf r}-{\bf r}')\delta(t-t')$ and $ p$ is the pressure that enforces the three-dimensional incompressibility constraint $\nabla\cdot{\bf v}=0$. In equilibrium, in addition to all active terms being $0$, ${D_m}/{\Gamma_m}={D}_v/{\eta}$. This completes the description of a chiral non-conserved order parameter in a momentum conserved system.

\section{Layered states}
\label{sec2}
In this section, we will consider layered states of both systems with conserved and non-conserved order parameters and derive their dynamical equations. We will first consider a layered state in a two-dimensional film and then in a three-dimensional system. 
\subsection{Layered states in a two-dimensional film}
\label{2dlayer}
We will start with the dynamics described in Sections \ref{2dH} and \ref{2dA} and derive the dynamical equations for a layered state -- a positionally ordered array of lines -- in two dimensions. We will first discuss this for the case of a conserved order parameter (Sec. \ref{2dH}) and then a non-conserved order parameter (Sec. \ref{2dA}). As expected, we will demonstrate that in both cases, the hydrodynamic theory we derive for the layered state will be the same.
\subsubsection{Two-dimensional layered state in a system with a conserved order parameter}
\label{2dconslayer}
We consider a layered state that can arise in a two-dimensional film that break up-down symmetry, with a chiral, conserved composition variable and derive its equation of motion starting from \eqref{psieq2d} and \eqref{vel2dpsi}. 
We will consider the effect of activity on a layered state of the composition field $\psi$ that may be realised \emph{in the absence of activity}, i.e, when $\lambda_1$, $\lambda_2$, $\omega_1$, $\omega_2$, $\zeta_H$ and $\bar{\zeta}_c$ in \eqref{acpsi2}, \eqref{ccpsi2} and \eqref{vel2dpsi} are set to $0$. The composition field $\psi$ may have mean value $\psi_0$ about which it has a periodic spatial modulation $\psi_1({\bf r})$ i.e, $\psi=\psi_0+\psi_1$.
The periodically modulated steady-state has no flow and minimises the standard Swift-Hohenberg free energy i.e., ${\delta  F [\psi]}/{\delta\psi}=0$ for 
\begin{equation}
\label{2dpsifenerg}
 F [\psi]= F [\psi_0]+\frac{\Upsilon}{2}\int d{\bf r}\left[-2 q_s^2(\nabla\psi_1)^2+(\nabla^2\psi_1)^2+\frac{\alpha}{2}(\psi_1)^2+\frac{\beta}{4}(\psi_1)^4\right],
\end{equation}
with $ q_s^{-1}$ being the periodicity of the layered state which is reached when the homogeneous state is destabilised for $\alpha<0$. Without any loss of generality, we assume that periodic modulation of $\psi$ is along $\hat{x}$ i.e., $\psi_1$ forms a state with a uniformly spaced array of lines whose normals are along $\hat{\bf z}$. This implies that the steady state $\psi_1$ is
\begin{equation}
\psi_1|_{s.s}=\psi_1^0[e^{i\phi_0}+e^{-i\phi_0}],
\end{equation}
where the amplitude $\psi_1^0=\sqrt{|\alpha|/\beta}$ and the phase is $\phi_0= q_s z$. We now consider the hydrodynamic fluctuations of $\psi_1$ about this \emph{passive} steady-state in the presence of active forces. The fluctuations of the \emph{amplitude} of $\psi_1$ are massive and relax to $\psi_1^0$ in a finite timescale. However, the phase fluctuations \emph{are} hydrodynamic. We therefore take 
\begin{equation}
\phi=\phi_0- q_s u(\SK{x},z,t)\equiv  q_s[z-u(\SK{x},z,t)]
\end{equation}
where $ u$ is the Goldstone mode of the the broken translational symmetry and denotes the displacement of the periodic array of layers from their steady state positions. Inserting 
\begin{equation}
\psi_1=\psi_1^0[e^{i\phi}+e^{-i\phi}]
\end{equation}
into \eqref{2dpsifenerg} we obtain a free energy purely in terms of $ u$
\begin{equation}
 F [\psi]= F [\psi_0]+2\Upsilon({\psi_1^{0}} q_s^2)^2\int\left[\{{\partial}_z  u-(1/2)(\nabla  u)^2\}^2+\mu^2(\nabla^2  u)^2\right],
\end{equation}
where $\mu\propto  q_s^{-1}$ \cite{GrinPel}. We now define $ B=(4 q_s^2\psi_1^{0})^2\Upsilon$ and $ K={(4 q_s^2\psi_1^{0}\mu)^2}\Upsilon$ to obtain the standard free energy for a layered state:
\begin{equation}
F [u]=\int\left[\frac{ B}{2}\left({\partial}_z  u-\frac{(\nabla  u)^2}{2}\right)^2+\frac{ K}{2}(\nabla^2  u)^2\right].
\end{equation}
The first term in the free energy is the compression modulus and involves the covariant strain
\begin{equation}
 E=\left({\partial}_z  u-\frac{(\nabla  u)^2}{2}\right).
\end{equation}
We now construct a dynamical equation for $ u$ from \eqref{psieq2d} and write the velocity equation from \eqref{vel2dpsi} in terms of $ u$. The phase gradient, which is along the normal to the layers, is $ q_s{\bf n}=\nabla\phi$ (note that ${\bf n}$ is \emph{not} a unit vector unlike in \cite{Tapan1}). The time evolution of $\psi$ reduces to 
\begin{equation}
\partial_t\psi=-i\psi_1^{0} q_s[e^{i\phi}-e^{-i\phi}]\partial_t  u.
\end{equation}
The velocity coupling from \eqref{ccpsi2} $\nabla\cdot[\omega_v\psi\nabla(\boldsymbol{\varepsilon}:\nabla{\bf v})]$ yields a term 
\begin{equation}
i\omega_v\psi_1^{0}[e^{i\phi}-e^{-i\phi}]\nabla\phi\cdot\nabla(\boldsymbol{\varepsilon}:\nabla{\bf v})=i\omega_v\psi_1^{0} q_s[e^{i\phi}-e^{-i\phi}]{\bf n}\cdot\nabla(\boldsymbol{\varepsilon}:\nabla{\bf v}),
\end{equation}
which enters the phase equation. Similarly, the achiral active terms \eqref{acpsi2} also contribute to the phase equation:
\begin{equation}
i\lambda_1\text{Im}[\nabla\cdot\{\psi\nabla\psi\nabla^2\psi\}]=-i\psi_1^{0^3}\lambda_1\nabla\phi\cdot\nabla(\nabla\phi)^2[e^{i\phi}-e^{-i\phi}]=2i\psi_1^{0^3} q_s^3\lambda_1{\bf n}\cdot\nabla E[e^{i\phi}-e^{-i\phi}]
\end{equation}
and
\begin{equation}
i\lambda_2\text{Im}[\nabla\cdot\{\psi\nabla(\nabla\psi)^2\}]=-i\psi_1^{0^3}\lambda_2[\nabla\phi\cdot\nabla(\nabla\phi)^2+2\nabla^2\phi(\nabla\phi)^2][e^{i\phi}-e^{-i\phi}]=2i\psi_1^{0^3} q_s^3\lambda_2[{\bf n}\cdot\nabla E-\nabla\cdot{\bf n}(1-2 E)][e^{i\phi}-e^{-i\phi}].
\end{equation}
The chiral active term, $\omega_2\nabla\cdot[\psi\boldsymbol{\varepsilon}\cdot\nabla(\nabla\psi)^2]=\omega_2\varepsilon_{ij}\partial_i\psi\partial_j(\partial_l\psi\partial_l\psi)$ has a term $3i\omega_2\psi_1^{0^3}\varepsilon_{ij}[e^{i\phi}-e^{-i\phi}]\partial_i\phi\partial_j(\partial_l\phi\partial_l\phi)$. Using
\begin{equation}
\partial_l\phi\partial_l\phi= q_s^2(\delta_{lx}-\partial_l u)(\delta_{lx}-\partial_l u)= q_s^2(1-2 E),
\end{equation}
we get
\begin{equation}
3i\omega_2\psi_1^{0^3}\varepsilon_{ij}[e^{i\phi}-e^{-i\phi}]\partial_i\phi\partial_j(\partial_l\phi\partial_l\phi)=3i\omega_2\psi_1^{0^3} q_s^3[e^{i\phi}-e^{-i\phi}]\varepsilon_{ij}{n}_i\partial_j(1-2 E)=-6i\omega_2\psi_1^{0^3} q_s^3[e^{i\phi}-e^{-i\phi}]\varepsilon_{ij}{n}_i\partial_j E.
\end{equation}
Similarly, treating the achiral currents and putting all of these together, the displacement field equation is 
\begin{equation}
\label{ueqn2d}
\partial_t u={\bf v}\cdot{\bf n}-\chi_v{\bf n}\cdot\nabla(\boldsymbol{\varepsilon}:\nabla{\bf v})+\Lambda_1{\bf n}\cdot\nabla E+\Lambda_2\nabla\cdot{\bf n}(1-2 E)-\chi (\boldsymbol{\varepsilon}:{\bf n}\nabla) E-{\Gamma}_u\frac{\delta  F [u]}{\delta  u}+{\xi}_u,
\end{equation}
where we have defined $\chi_v=\omega_v$, $\Lambda_1=-2\psi_1^{0^2} q_s^2(\lambda_1+\lambda_2)$, $\Lambda_2=2\psi_1^{0^2} q_s^2\lambda_2$, $\chi =\psi_1^{0^2} q_s^2(2\omega_1-6\omega_2)$, ${\Gamma}_u=-{M} q_s^2$ and $\langle{\xi}_u({\bf r},t){\xi}_u({\bf r}',t')\rangle=2{D}_u\delta({\bf r}-{\bf r}')\delta(t-t')$. We now consider the constitutive equation for the Stokesian velocity field. For this, we  consider the term
\begin{multline}
\label{strs2du}
(\nabla\psi\nabla\psi)^{ST}=2\psi_1^{0^2}[\nabla\phi\nabla\phi-{\bsf I}(\nabla\phi)^2/2]=2\psi_1^{0^2} q_s^2[{\bf n}{\bf n}-(1/2){\bsf I}{\bf n}\cdot{\bf n}]\\=2\psi_1^{0^2} q_s^2[{\bf n}{\bf n}+( E-1/2){\bsf I}]=2\psi_1^{0^2} q_s^2[(\hat{\bf z}-\nabla  u)(\hat{\bf z}-\nabla  u)+{\bsf I}\{\partial_z  u-(\nabla  u)^2/2-1/2\}]\\=2\psi_1^{0^2} q_s^2\begin{pmatrix}-(1/2)+\partial_z  u-(\partial_z u)^2/2+(\partial_x u)^2/2 & -(1-\partial_z  u)\partial_x  u \\-(1-\partial_z  u)\partial_x  u & (1/2)-{\partial}_z  u+(\partial_z u)^2/2-(\partial_x u)^2/2\end{pmatrix}=-2\psi_1^{0^2} q_s^2\boldsymbol{w}.
\end{multline}
With this, \eqref{vel2dpsi} becomes
\begin{equation}
\label{vel2du}
\eta{\nabla}^2{\bf v}={\bf n}\frac{\delta F [ u]}{\delta u}+\nabla p+\nabla\cdot[\zeta {\bsf I}-\zeta _{c}\boldsymbol{\varepsilon}]\cdot\boldsymbol{w}+\chi_v(\boldsymbol{\varepsilon}\cdot\nabla)\nabla\cdot\left[{\bf n}\frac{\delta  F [ u]}{\delta u}\right]+{\boldsymbol{\xi}}_v,
\end{equation}
where $\zeta =2\psi_1^{0^2} q_s^2\zeta_H$ and $\zeta_c=-2\psi_1^{0^2} q_s^2\bar{\zeta}_c$. Eqs. \eqref{ueqn2d} and \eqref{vel2du} constitute a complete description of a layered state formed by chiral components in a two-dimensional thin film.

\subsubsection{Two-dimensional layered state in a system with a non-conserved order parameter}
\label{2dlaynoncons}
We will demonstrate that (unsurprisingly) the dynamics of a layered state formed by a non-conserved order parameter (Sec. \ref{2dA}) is eactly equivalent to that formed by a conserved order parameter. We will closely follow the path outlined in the last section and consider the effect of activity on a layered state of the field $ m $ in the absence of activity. We will assume that $ m $ has a mean value $ m _0$ about which it has periodic spatial modulations i.e. $ m = m _0+ m _1$. As in the last section we take the free energy to be 
\begin{equation}
\label{2dmfenerg}
 F [ m ]= F [ m _0]+\frac{\Upsilon}{2}\int d{\bf r}\left[-2 q_s^2(\nabla m _1)^2+(\nabla^2 m _1)^2+\frac{\alpha}{2}( m _1)^2+\frac{\beta}{4}( m _1)^4\right],
\end{equation}
with $ q_s^{-1}$ being the periodicity of the layered state which is reached when the homogeneous state is destabilised for $\alpha<0$. As earlier, we assume that periodic modulation of $m$ is along $\hat{\bf z}$. The steady state $ m _1$ is
\begin{equation}
 m _1|_{s.s}= m _1^0[e^{i\phi_0}+e^{-i\phi_0}]
\end{equation}
where the amplitude $ m _1^0=\sqrt{|\alpha|/\beta}$ and the phase is $\phi_0= q_s x$. Considering the soft phase fluctuations,
\begin{equation}
\phi=\phi_0-q_s u(y,z,t)\equiv  q_s[z-u(y,z,t)],
\end{equation}
where $u$ is the Goldstone mode of the the broken translational symmetry and denotes the displacement of the periodic array of layers from their steady state positions, we get
\begin{equation}
 m _1= m _1^0[e^{i\phi}+e^{-i\phi}].
\end{equation}
Inserting this into \eqref{2dmfenerg} we obtain a free energy purely in terms of $ u$:
\begin{equation}
 F [ m ]= F [\psi_0]+2\Upsilon({ m _1^{0}} q_s^2)^2\int\left[\{{\partial}_z  u-(1/2)(\nabla  u)^2\}^2+\mu^2(\nabla^2  u)^2\right],
\end{equation}
where $\mu\propto  q_s^{-1}$ \cite{GrinPel}. We now define $ B=(4 q_s^2 m _1^{0})^2\Upsilon$ and $ K={(4 q_s^2 m _1^{0}\mu)^2}\Upsilon$ to obtain the standard free energy for a layered state:
\begin{equation}
 F [u]=\int\left[\frac{ B}{2}E^2+\frac{ K}{2}(\nabla^2  u)^2\right].
\end{equation}
Writing the slow time evolution of $ m $ as
\begin{equation}
\partial_t m =-i m _1^{0} q_s[e^{i\phi}-e^{-i\phi}]\partial_t  u,
\end{equation}
and following the arguments of the last section we arrive at the equations of motion of $ u$ from \eqref{eqm2d} and the constitutive equation for ${\bf v}$, depending on $u$ from \eqref{vel2dm}:
\begin{equation}
\label{ueqn2dm}
\partial_t u={\bf v}\cdot{\bf n}-\chi_v{\bf n}\cdot\nabla(\boldsymbol{\varepsilon}:\nabla{\bf v})+\Lambda_1{\bf n}\cdot\nabla E+\Lambda_2\nabla\cdot{\bf n}(1-2 E)-\chi (\boldsymbol{\varepsilon}:{\bf n}\nabla) E-{\Gamma}_u\frac{\delta  F [u]}{\delta  u}+{\xi}_u,
\end{equation}
where $\chi_v=\omega_{v2}$, $\Lambda_1=-2 m _1^{0^2} q_s^2(\lambda_3+3\lambda_4+\lambda_6)$, $\Lambda_2=2 m _1^{0^2} q_s^2(-3\lambda_2+\lambda_3+2\lambda_4+2\lambda_5+\lambda_6)$, $\chi = m _1^{0^2} q_s^2(2\zeta _{m c1}-6\zeta _{m c2})$ and ${\Gamma}_u\propto\Gamma_m$. The velocity field equation is
\begin{equation}
\label{vel2dum}
\eta{\nabla}^2{\bf v}={\bf n}\frac{\delta F [ u]}{\delta u}+\nabla p+\nabla\cdot[\zeta {\bsf I}-\zeta _{c}\boldsymbol{\varepsilon}]\cdot\boldsymbol{w}+\chi_v(\boldsymbol{\varepsilon}\cdot\nabla)\nabla\cdot\left[{\bf n}\frac{\delta  F [ u]}{\delta u}\right]+{\boldsymbol{\xi}}_v,
\end{equation}
where $\zeta =2 m _1^{0^2} q_s^2\zeta_H$ and $\zeta_c=-2 m _1^{0^2} q_s^2\bar{\zeta}_c$. These equations are exactly the same as the ones derived in Sec. \ref{2dconslayer} with the difference being hidden in the relation between the phenomenological coefficients of the layered state and those introduced in Sec.  \ref{2dH} and  \ref{2dA}.

\subsection{Layered states in three-dimensional systems}
\label{3dlayer}
In this section, we will start with the dynamics described in Sections \ref{3dH} and \ref{3dA} and derive the dynamical equations for a layered state in three dimensions. We will first discuss this for the case of a conserved order parameter (Sec. \ref{3dH}) and then a non-conserved order parameter (Sec. \ref{3dA}). As expected, we will demonstrate that in both cases, the hydrodynamic theory we derive for the layered state are the same.
\subsubsection{Three-dimensional layered state in a system with a conserved order parameter}
\label{3dconslayer}
In this section, we consider a layered state that can arise in a three-dimensional system, with a chiral, conserved composition variable and derive its equation of motion starting from \eqref{psieq3d} and \eqref{vel3dpsi}. 
As earlier, we will consider the effect of activity on a layered state of the composition field ${\psi}$ that may be realised \emph{in the absence of activity}. The composition field ${\psi}$ may have mean value ${\psi}_0$ about which it has a periodic spatial modulation ${\psi}_1({\bf r})$ i.e, ${\psi}={\psi}_0+{\psi}_1$.
The periodically modulated steady-state has no flow and minimises free energy
\begin{equation}
\label{3dpsifenerg}
{F}[{\psi}]={F}[{\psi}_0]+\frac{{\Upsilon}}{2}\int d{\bf r}\left[-2{q}_s^2({\nabla}{\psi}_1)^2+({\nabla}^2{\psi}_1)^2+\frac{{\alpha}}{2}({\psi}_1)^2+\frac{{\beta}}{4}({\psi}_1)^4\right],
\end{equation}
with ${q}_s^{-1}$ being the periodicity of the layered state which is reached when the homogeneous state is destabilised for ${\alpha}<0$. We take the periodic modulation of ${\psi}$ to be along $\hat{z}$ i.e., ${\psi}_1$ forms a state with a uniformly spaced array of layers whose normals are along $\hat{z}$. This implies that the steady state ${\psi}_1$ is
\begin{equation}
{\psi}_1|_{s.s}={\psi}_1^0[e^{i{\phi}_0}+e^{-i{\phi}_0}],
\end{equation}
where the amplitude ${\psi}_1^0=\sqrt{|\alpha|/\beta}$ and the phase is ${\phi}_0={q}_s z$. We now consider the hydrodynamic fluctuations of ${\psi}_1$ about this \emph{passive} steady-state in the presence of active forces. The fluctuations of the \emph{amplitude} of ${\psi}_1$ are massive and relax to ${\psi}_1^0$ in a finite timescale. However, the phase fluctuations \emph{are} hydrodynamic. We therefore take 
\begin{equation}
{\phi}={\phi}_0-q_s{u}(x,y,z,t)\equiv {q}_s[z-u(x,y,z,t)],
\end{equation}
where ${u}$ is the Goldstone mode of the the broken translational symmetry and denotes the displacement of the periodic array of layers from their steady state positions. Inserting 
\begin{equation}
{\psi}_1={\psi}_1^0[e^{i{\phi}}+e^{-i{\phi}}]
\end{equation}
into \eqref{3dpsifenerg} we obtain a free energy purely in terms of ${u}$,
\begin{equation}
{F}[{\psi}]={F}[{\psi}_0]+2{\Upsilon}({{\psi}_1^{0}}q_s^2)^2\int\left[\{\SK{{\partial}_z {u}}-(1/2)({\nabla}  u)^2\}^2+{\mu}^2({\nabla}^2 {u})^2\right],
\end{equation}
where $\mu\propto q_s^{-1}$ \cite{GrinPel}. We now define ${B}=(4q_s^2{\psi}_1^{0})^2{\Upsilon}$ and ${K}={(4q_s^2{\psi}_1^{0}{\mu})^2}{\Upsilon}$ to obtain the standard free energy for a layered state
\begin{equation}
{F}[u]=\int\left[\frac{{B}}{2}\left({\partial}_z {u}-\frac{({\nabla} {u})^2}{2}\right)^2+\frac{{K}}{2}({\nabla}^2 {u})^2\right].
\end{equation}
The first term in the free energy is the compression modulus and involves the covariant strain
\begin{equation}
{E}=\left({\partial}_z {u}-\frac{({\nabla} {u})^2}{2}\right).
\end{equation}
 The normal to the layers is ${q}_s{{\bf n}}={\nabla}{\phi}$ (note that ${{\bf n}}$ is \emph{not} a unit vector unlike in \cite{Tapan1}). The time derivative of $\psi$ yields
\begin{equation}
\partial_t{\psi}=-i{\psi}_1^{0}{q}_s[e^{i{\phi}}-e^{-i{\phi}}]\partial_t {u}.
\end{equation}
We construct a dynamical equation for ${u}$ from \eqref{psieq3d} and write the velocity equation \eqref{vel3dpsi} in terms of ${u}$:
\begin{equation}
\label{ueqn3d}
\partial_t{u}={{\bf v}}\cdot{{\bf n}}+ C_v{{\bf n}}\cdot[{\nabla}^2(\nabla\times{\bf v})]+\Lambda_1{{\bf n}}\cdot\nabla{E}+\Lambda_2{\nabla}\cdot{{\bf n}}(1-2{E})-{\Gamma}_u\frac{\delta {F}[u]}{\delta {u}}+{\xi}_u,
\end{equation}
where $ C_v=-\Omega_v$, $\Lambda_1=-2{\psi}_1^{0^2}{q}_s^2(\lambda_1+\lambda_2)$, $\Lambda_2=2{\psi}_1^{0^2}{q}_s^2\lambda_2$, ${\Gamma}_u=-{M}{q}_s^2$ and $\langle{\xi}_u({\bf r},t){\xi}_u({\bf r}',t')\rangle=2{D}_u\delta({\bf r}-{\bf r}')\delta(t-t')$. Using
\begin{multline}
\label{strs3du}
(\partial_i{\psi}\partial_j{\psi})=2{\psi}_1^{0^2}[\partial_i\phi\partial_j\phi]=2{\psi}_1^{0^2}{q}_s^2[n_in_j]=2{\psi}_1^{0^2}{q}_s^2[(\delta_{iz}-\partial_iu)(\delta_{jz}-\partial_ju)]\\=2{\psi}_1^{0^2}{q}_s^2\begin{pmatrix}({\partial}_x {u})^2 & \partial_xu\partial_y u & -(1-\partial_z u)\partial_x u \\ \partial_xu\partial_y u & (\partial_y u)^2 & -(1-\partial_z u)\partial_y u\\ -(1-\partial_z u)\partial_x u & -(1-\partial_z u)\partial_y u & (1-\partial_z u)^2\end{pmatrix}=-2{\psi}_1^{0^2}{q}_s^2 w_{ij}
\end{multline}
we rewrite \eqref{vel3dpsi} as 
\begin{equation}
\label{vel3du}
\eta\nabla^2{v}_i=n_i\frac{\delta {F}[u]}{\delta {u}}+\partial_i p+\partial_j[{\zeta} w_{ij}+{\zeta}_{c}\partial_l(\epsilon_{ijk} w_{kl})]+ C_v\epsilon_{ijk}\partial_j\partial_l\partial_l\left[n_k\frac{\delta F[u]}{\delta u}\right]+\xi_{v_i},
\end{equation}
where $\zeta=2{\psi}_1^{0^2}{q}_s^2\zeta_H$ and $ z_c=2{\psi}_1^{0^2}{q}_s^2 \bar{z}_c $. Eqs. \eqref{ueqn3d} and \eqref{vel3du} constitute the complete description of an active, chiral layered state.

\subsubsection{Three-dimensional layered state in a system with a non-conserved order parameter}
\label{3dlaynoncons}
In this section, we consider a layered state that can arise in a three-dimensional system, with a chiral, non-conserved composition variable and derive its equation of motion starting from \eqref{3dmeq} and \eqref{vel3dm}. 
As earlier, we will consider the effect of activity on a layered state of the order parameter field ${m}$ that may be realised \emph{in the absence of activity}. The order parameter field ${m}$ may have mean value ${m}_0$ about which it has a periodic spatial modulation ${m}_1({\bf r})$ i.e, ${m}={m}_0+{m}_1$.
The periodically modulated steady-state has no flow and minimises free energy
\begin{equation}
\label{3dmfenerg}
{F}[{m}]={F}[{m}_0]+\frac{{\Upsilon}}{2}\int d{\bf r}\left[-2{q}_s^2({\nabla}{m}_1)^2+({\nabla}^2{m}_1)^2+\frac{{\alpha}}{2}({m}_1)^2+\frac{{\beta}}{4}({m}_1)^4\right],
\end{equation}
with ${q}_s^{-1}$ being the periodicity of the layered state which is reached when the homogeneous state is destabilised for ${\alpha}<0$. We assume that periodic modulation of ${m}$ is along $\hat{z}$ i.e., ${\psi}_1$ forms a state with a uniformly spaced array of layers whose normals are along $\hat{z}$. This implies that the steady state ${m}_1$ is
\begin{equation}
{m}_1|_{s.s}={m}_1^0[e^{i{\phi}_0}+e^{-i{\phi}_0}],
\end{equation}
where the amplitude ${m}_1^0=\sqrt{|\alpha|/\beta}$ and the phase is ${\phi}_0={q}_s z$. Considering the fluctuations
\begin{equation}
{\phi}={\phi}_0-q_s{u}(x,y,z,t)\equiv {q}_s[z-u(x,y,z,t)],
\end{equation}
where ${u}$ is the Goldstone mode of the the broken translational symmetry and denotes the displacement of the periodic array of layers from their steady state positions, and inserting 
\begin{equation}
{m}_1={m}_1^0[e^{i{\phi}}+e^{-i{\phi}}]
\end{equation}
into \eqref{3dmfenerg} we obtain a free energy purely in terms of ${u}$:
\begin{equation}
{F}[{m}]={F}[{m}_0]+2{\Upsilon}({{m}_1^{0}}q_s^2)^2\int\left[\{{\partial}_x {u}-(1/2)({\nabla}  u)^2\}^2+{\mu}^2({\nabla}^2 {u})^2\right],
\end{equation}
where $\mu\propto q_s^{-1}$ \cite{GrinPel}. We define ${B}=(2q_s^2{m}_1^{0})^2{\Upsilon}$ and ${K}={(2q_s^2{m}_1^{0}{\mu})^2}{\Upsilon}$ to obtain the standard free energy for a layered state:
\begin{equation}
{F}[u]=\int\left[\frac{{B}}{2}\left({\partial}_z {u}-\frac{({\nabla} {u})^2}{2}\right)^2+\frac{{K}}{2}({\nabla}^2 {u})^2\right]=\int\left[\frac{{B}}{2}E^2+\frac{{K}}{2}({\nabla}^2 {u})^2\right].
\end{equation}
The normal to the layers is ${q}_s{{\bf n}}={\nabla}{\phi}$. The time derivative of $m$ yields
\begin{equation}
\partial_t{m}=-i{m}_1^{0}{q}_s[e^{i{\phi}}-e^{-i{\phi}}]\partial_t {u}.
\end{equation}
We now construct a dynamical equation for ${u}$ from \eqref{3dmeq} and the velocity equation from \eqref{vel3dm} in terms of ${u}$:
\begin{equation}
\label{ueqn3d2}
\partial_t{u}={{\bf v}}\cdot{{\bf n}}+ C_v{{\bf n}}\cdot[{\nabla}^2(\nabla\times{\bf v})]+\Lambda_1{{\bf n}}\cdot\nabla{E}+\Lambda_2{\nabla}\cdot{{\bf n}}(1-2{E})-{\Gamma}_u\frac{\delta {F}[u]}{\delta {u}}+{\xi}_u,
\end{equation}
where $ C_v=-\Omega_v$, $\Lambda_1=-2{m}_1^{0^2}{q}_s^2(\lambda_3+3\lambda_4+\lambda_6)$, $\Lambda_2=2{m}_1^{0^2}{q}_s^2(-3\lambda_2+\lambda_3+2\lambda_4+2\lambda_5+\lambda_6)$, ${\Gamma}_u\propto\Gamma_m$, and
\begin{equation}
\label{vel3du2}
\eta\nabla^2{v}_i=n_i\frac{\delta {F}[u]}{\delta {u}}+\partial_i p+\partial_j[{\zeta} w_{ij}+{\zeta}_{c}\partial_l(\epsilon_{ijk} w_{kl})]+ C_v\epsilon_{ijk}\partial_j\partial_l\partial_l\left[n_k\frac{\delta F[u]}{\delta u}\right]+\xi_{v_i},
\end{equation}
where $\zeta=2{m}_1^{0^2}{q}_s^2\zeta_H$ and $ z_c=2{m}_1^{0^2}{q}_s^2\bar{z}_c$. As expected, these are exactly the same equations as those obtained in the last section.

\section{Equivalence of active stress and external stress}
\label{Eqastrestr}
In this section, we will demonstrate that the influence of the achiral active stress in a layered state in an incompressible system is equivalent to an \emph{externally} imposed stress. To demonstrate this, we introduce an external stress through the additional term in the free energy
\begin{equation}
F[u]\to F[u]+F^{ext} [u]=F[u]+\int\sigma_0 E.
\end{equation}
The force in the momentum density equation due to the external stress is,
\begin{equation}
\SK{f_{i}^{ext}}=-\frac{\delta F^{ext}}{\delta u}(\delta_{iz}-\partial_iu).
\end{equation}
\begin{equation}
\frac{\delta F^{ext}}{\delta u}=-\sigma_0\partial_j\frac{\partial E}{\partial \partial_j u}=-\sigma_0\partial_j(\delta_{jz}-\partial_ju).
\end{equation}
This implies that
\begin{multline}
\SK{f_{i}^{ext}}=-\frac{\delta F^{ext}}{\delta u}(\delta_{iz}-\partial_iu)=\sigma_0\partial_j\frac{\partial E}{\partial \partial_j u}(\delta_{iz}-\partial_iu)=\sigma_0\left(\frac{\partial E}{\partial \partial_i u}\right)\partial_j\left(\frac{\partial E}{\partial \partial_j u}\right)\\=\sigma_0\partial_j\left[\left(\frac{\partial E}{\partial \partial_i u}\right)\left(\frac{\partial E}{\partial \partial_j u}\right)\right]-\sigma_0\left(\frac{\partial E}{\partial \partial_j u}\right)\partial_j\left(\frac{\partial E}{\partial \partial_i u}\right).
\end{multline}
Using
\begin{equation}
\left(\frac{\partial E}{\partial \nabla u}\right)\cdot\nabla\left(\frac{\partial E}{\partial \nabla u}\right)=-\nabla[\partial_z u-(1/2)(\nabla u)^2]=-\nabla E,
\end{equation}
we get
\begin{equation}
{\bf f}^{ext}=\sigma_0\nabla\cdot\left[\left(\frac{\partial E}{\partial \nabla u}\right)\left(\frac{\partial E}{\partial \nabla u}\right)+E{\bsf I}\right]=-\nabla\cdot\boldsymbol{\sigma}^{ext}.
\end{equation}
The isotropic part of $\boldsymbol{\sigma}^{ext}$ is, of course, unimportant in an incompressible system since it can be absorbed in the pressure and 
\begin{equation}
\left(\frac{\partial E}{\partial \nabla u}\right)\left(\frac{\partial E}{\partial \nabla u}\right)=
\begin{pmatrix}({\partial}_x {u})^2 & \partial_xu\partial_y u & -(1-\partial_z u)\partial_x u \\ \partial_xu\partial_y u & (\partial_y u)^2 & -(1-\partial_z u)\partial_y u\\ -(1-\partial_z u)\partial_x u & -(1-\partial_z u)\partial_y u & (1-\partial_z u)^2\end{pmatrix}.
\end{equation}
This is \emph{exactly} equivalent to the \emph{active} stress in \eqref{strs3du}. This implies that the active stress in \eqref{vel3du} is indistinguishable from an \emph{external} stress, with $\SK{\sigma_0=\zeta}$, up to an irrelevant, isotropic piece. This further implies that \emph{all} effects of the achiral active stress in a layered system can be eliminated by exerting an external stress such that $\sigma_0+\zeta=0$ -- such an externally stressed \emph{active} layered state would be indistinguishable from a \emph{passive} layered system \emph{without} an external stress.

Exactly the same calculation can be carried out for a two-dimensional layered state. In this case,
\begin{equation}
\boldsymbol{\sigma}^{ext}=-\sigma_0\left[\left(\frac{\partial  E}{\partial \nabla  u}\right)\left(\frac{\partial  E}{\partial \nabla  u}\right)+ E{\bsf I}\right]=-\sigma_0\begin{pmatrix} \partial_z  u-(\partial_z u)^2/2+(\partial_x u)^2/2& -(1-\partial_z u)\partial_x  u \\-(1-\partial_z  u)\partial_x  u & 1-{\partial}_z  u+(\partial_z u)^2/2-(\partial_x u)^2/2 \end{pmatrix}.
\end{equation}
This differs from the two-dimensional active stress $\zeta\boldsymbol{w}$ in \eqref{strs2du} by an isotropic piece $\propto -1/2{\bsf I}$, when $\sigma_0=\zeta$ which is \emph{independent} of $ u$. Therefore, in both three and two dimensions, an achiral active stress can be compensated by an external imposed stress.

Of course, from \eqref{ueqn3d} and \eqref{vel3du} we find that the free energy enters both in the $u$ equation, with a permeative coefficient $\Gamma_u$, and the velocity equation. Therefore, in passive, externally stressed layered systems, $\sigma_0$ also enters \eqref{ueqn3d} through $\partial_tu\propto-\Gamma_u\sigma_0\nabla^2 u$. There is no obligation for a term $\propto\nabla^2 u$ in the $\dot{u}$ equation to have the same coefficient as the active stress $\zeta$. Therefore, strictly speaking, the equivalence between an externally stressed layered state and one with an achiral active stress is \emph{only} valid for \emph{impermeable} systems i.e., one in which $\Lambda_1=\Lambda_2=\Gamma_u=0$.

\section{Equivalence of different active models for layered states}
\label{equiallstr}
In this paper, we have introduced activity via ``active'' terms in the equations of motion (see, for instance, \eqref{psieq3d} and \eqref{vel3dpsi} in Sec. \ref{3dH} or the corresponding equations in Sec. \ref{2dH}, \ref{2dA}, \ref{3dA}). However, the \emph{same} energy function $F[\psi]$ appears in both \eqref{psieq3d} and \eqref{vel3dpsi}. That is neither necessary nor inevitable in active systems. We will now show that taking two different free energies, $F_1[\psi]$ in \eqref{psieq3d} and $F_2[\psi]$ \eqref{vel3dpsi} would not have led to any qualitatively new effect in the layered state. Since we are interested in the phase dynamics, we assume both the free energies have the same preferred $\psi_1^0$. We take
\begin{equation}
\label{fenergt1}
{F}_2[{\psi_1}]=\frac{{\Upsilon_1}}{2}\int d{\bf r}\left[-2{q}_1^2({\nabla}{\psi}_1)^2+({\nabla}^2{\psi}_1)^2+\frac{{\alpha}}{2}{\psi}_1^2+\frac{{\beta}}{4}{\psi}_1^4\right]
\end{equation}
and
\begin{equation}
\label{fenergt2}
{F}_2[{\psi_1}]=\frac{{\Upsilon_2}}{2}\int d{\bf r}\left[-2{q}_2^2({\nabla}{\psi}_1)^2+({\nabla}^2{\psi}_1)^2+\frac{{\alpha}}{2}{\psi}^2+\frac{{\beta}}{4}{\psi}_1^4\right]
\end{equation}
to be the two free energies. Further, as discussed earlier, the wavevector of the layered pattern in an active system need not be (and, in general, will not be) the one that would be selected in a passive system. In particular, this means that the selected wavelength of the pattern 
\begin{equation}
\psi_1=\psi_1^0[e^{i\phi_0}+e^{-i\phi_0}],
\end{equation}
where $\psi_1^0=\sqrt{|\alpha|/\beta}$ and $\phi_0=q_s z$, need not be equal to either $q_1$ or $q_2$ i.e. $q_s\neq q_1\neq q_2$, in general. Inserting the form of $\psi_1$ in \eqref{fenergt1} and \eqref{fenergt2}, we obtain
\begin{equation}
F_1[u]=2\Upsilon_1(\psi^0_1q_s^2)^2\int\left[\frac{q_1^2-q_s^2}{q_s^2}\{\partial_z u-(1/2)(\nabla u)^2\}+\{\partial_z u-(1/2)(\nabla u)^2\}^2+\mu_1^2(\nabla^2 u)^2\right]
\end{equation}
and
\begin{equation}
F_2[u]=2\Upsilon_2(\psi^0_1q_s^2)^2\int\left[\frac{{q}_2^2-q_s^2}{q_s^2}\{\partial_z u-(1/2)(\nabla u)^2\}+\{\partial_z u-(1/2)(\nabla u)^2\}^2+\mu_2^2(\nabla^2 u)^2\right].
\end{equation}
Writing $\sigma_1=2\Upsilon_1(\psi^0_1q_s)^2(q_1^2-q_s^2)$, $B_1=4\Upsilon_1(\psi^0_1q_s^2)^2$, $K_1=4\Upsilon_1(\psi^0_1q_s^2)^2\mu_1^2$, $\sigma_2=2\Upsilon_2(\psi^0_1q_s)^2(q_2^2-q_s^2)$, $B_2=4\Upsilon_2(\psi^0_1q_s^2)^2$, $K_2=4\Upsilon_2(\psi^0_1q_s^2)^2\mu_2^2$, these free energies can be written in the form
\begin{equation}
F_1[u]=\frac{1}{2}\int 2\sigma_1 E+B_1E^2+K_1(\nabla^2u)^2
\end{equation}
and 
\begin{equation}
F_2[u]=\frac{1}{2}\int 2\sigma_2 E+B_2E^2+K_2(\nabla^2u)^2.
\end{equation}
The dominant effect of the two free energy picture arises from the effective ``excess'' stress terms $\sigma_1$ and $\sigma_2$. As discussed in the last section, $\sigma_2$ essentially amounts to a shift of the achiral active stress in \eqref{vel3du}. Similarly, $\sigma_1$ would induce a permeative term with $\partial_u \propto \Gamma_u\sigma_1\nabla^2 u$ in \eqref{ueqn3d}. Thus, since \eqref{ueqn3d} \emph{already} contains active permeative terms $\propto \nabla^2 u$, neither $\sigma_1$ nor $\sigma_2$ leads to any qualitatively new dynamics beyond those captured by \eqref{ueqn3d} and \eqref{vel3du}. Similarly, since \eqref{ueqn3d} already contains active permeative terms proportional to ${\bf n}\cdot\nabla E$, the difference between $B_1$ and $B_2$ also doesn't lead to any qualitatively new effect (beyond a trivial shift of an active permeative coefficient). The same conclusion can be drawn about the bending free energies. This implies that i. no qualitatively new phenomenon can be expected if, in addition to the active terms we have introduced, we also allowed for the possibility of distinct free energies in different equations and ii. though, in general, the ordering wavevector in an active system is different from its passive counterpart, this difference only leads to a simple shift in the achiral active stress and implies that our conclusions, obtained by examining the effect of activity on a \emph{passive} steady-state, is valid more generally.

Now we discuss a second possible distinction between the model we introduce here and other models of active layered states. In this work, we construct active stresses out of the dyadic $\nabla\phi\nabla\phi$. However, other works on active layered systems (for instance, \cite{Tapan1, Tapan2}) introduce an active stress proportional to ${\bf mm}$ where ${\bf m}$, distinct from our ${\bf n}=(1/q_s)\nabla\phi$, is the \emph{unit} normal ${\bf m}= \nabla\phi/|\nabla\phi|$. We now show that this doesn't lead to any physics unaccounted for within our treatment. To demonstrate this, we first calculate the form of the ``director active stress'':
\begin{equation}
\label{dirstrs}
{\sigma}^{dir}_{ij}=\zeta{\bf mm}=\zeta\frac{1}{|\nabla\phi|^2}\partial_i\phi\partial_j\phi=\zeta\frac{1}{1-2E}(\delta_{iz}-\partial_iu)(\delta_{jz}-\partial_ju)\approx \zeta(1+2E)(\delta_{iz}-\partial_iu)(\delta_{jz}-\partial_ju),
\end{equation}
where we have used the relation
\begin{equation}
|\nabla\phi|^2=q_s^2[(1-\partial_zu)^2+(\nabla_\perp u)^2]=q_s^2(1+(\nabla u)^2-2\partial_z u)=q_s^2(1-2E),
\end{equation}
and where the last approximate equality in \eqref{dirstrs} is for small layer deformations. This demonstrates that at small deformations, the director form of the active stress leads to an extra term $\propto 2\zeta E(\delta_{iz}-\partial_iu)(\delta_{jz}-\partial_ju)$. We will now show that this can be absorbed into a redefinition of the free energy in the \emph{velocity} equation (but without any equivalent redefinition in the $\dot{u}$ equation, for permeative systems). To demonstrate this, we closely follow \eqref{Eqastrestr} to calculate the stress due to the compressive free energy:
\begin{equation}
\label{elasstrs}
\sigma_{ij}^{elas}=-BE\left(\frac{\partial E}{\partial \partial_i u}\right)\left(\frac{\partial E}{\partial \partial_j u}\right)-\frac{BE^2}{2}\delta_{ij}=-BE(\delta_{iz}-\partial_iu)(\delta_{jz}-\partial_ju)-\frac{BE^2}{2}\delta_{ij}.
\end{equation}
Thus, up to the isotropic piece $\propto BE^2/2{\bsf I}$ which can be absorbed into a redefinition of the pressure, the form of \eqref{elasstrs} is \emph{exactly} equivalent to the extra term $\propto 2\zeta E(\delta_{iz}-\partial_iu)(\delta_{jz}-\partial_ju)$ (for small deformations) that using the director form of the active stress leads to. This part of the ``active stress'' can therefore be absorbed into a redefinition of the compression modulus of the layered sate, but \emph{only} in the velocity equation. 
Crucially, therefore, in a layered state in which \emph{permeation} is allowed, this amounts to a description with \emph{two} distinct free energies. However, as discussed above, such a discussion is \emph{not} qualitatively distinct from one which starts from \eqref{ueqn3d} and \eqref{vel3du} -- the ``different free energies'' can be reconciled with our description with a single free energy via a \emph{further} shift of an active permeative coefficient. This implies that, at least for small deformations, our description and one with an active stress $\propto {\bf mm}$ are equivalent up to a redefinition of phenomenological coefficients.

\section{Linearised theory of chiral layered states}
\label{linchisec}
We will now consider the linearised theory of chiral layered states. We will first consider a layered state in a two-dimensional thin film (see Sec. \ref{2dlayer}) and then a three-dimensional layered state (see Sec. \ref{3dlayer}). We first consider a two-dimensional layered state discussed in Sec. \ref{2dlayer}
\subsection{Linear theory of two-dimensional, chiral layered states}
\label{linchisec2d}
In this section we will examine the theory of linearised fluctuations about a layered state implied by \eqref{ueqn2d} and \eqref{vel2du} in Sec. \ref{2dconslayer} or equivalently, \eqref{ueqn2dm} and \eqref{vel2dum} in Sec. \ref{2dlaynoncons}. The linearised equations of motion are
\begin{equation}
\label{linubar}
\partial_t u= v_z+\chi_v\partial_z(\partial_z v_x-\partial_x v_z)+\Lambda_1\partial_z^2 u-\Lambda_2\nabla^2 u+\chi \partial_z\partial_x u+{\Gamma}_u B\partial_z^2 u-{\Gamma}_u K\nabla^4 u+\xi_u,
\end{equation}
\begin{equation}
\label{linvybar}
\eta\nabla^2 v_x=\partial_x p-\zeta_c\nabla^2 u-\chi_v\partial_z^2( B\partial_z^2 u- K\nabla^4 u)+\xi_{vx}
\end{equation}
and
\begin{equation}
\label{linvxbar}
\eta\nabla^2 v_z=-( B\partial_z^2 u- K\nabla^4 u)+\partial_z p+\zeta \nabla^2 u+\chi_v\partial_x\partial_z( B\partial_z^2 u- K\nabla^4 u)+\xi_{vz}
\end{equation}
Before analysing these equations of motion, a few comments are in order. The chiral active stress implies that a curvature, in addition to exerting a force along the normal to the layers (due to the achiral active force), can also exert a force \emph{transverse} to the layers. This crucially requires the breaking of up-down symmetry.
The term with coefficient $\chi_v$ in \eqref{linubar} implies that a $z$ gradient of the two-dimensional vorticity leads to a translation of the layers. This is a chiral velocity coupling that is present even in equilibrium. This implies that creating a Poiseuille profile of the velocity \emph{transverse} to the layers leads to a motion of the layers. That is, unlike in achiral layered states, a flow transverse to the layers leads to a drift of the layers in chiral layered states. The chiral, active permeative term with the coefficient $\chi $ that \emph{tilting} a configuration of the layers with a uniform gradient along the layer normal direction leads to a drift of the layers. We now embark on an analysis of the linearied dynamics implied by \eqref{linubar}, \eqref{linvxbar} and \eqref{linvybar}. Using incompressibility to eliminate the pressure and solve \eqref{linvxbar} and \eqref{linvybar} in the Fourier space and inserting into the Fourier-transformed version of \eqref{linubar}, we obtain to $\mathcal{O}(q^0)$,
\begin{equation}
\label{linstab2d}
-i\omega  u=-\frac{1}{\eta q ^4}[\{ Bq_z^2q_x^2-(\zeta q_x^2+\zeta_cq_zq_x)q^2\} u+q_x^2\xi_{vz}-q_zq_x\xi_{vx}],
\end{equation}
where the noise $\xi_u$ originally in the $\partial_t  u$ equation is subdominant to the noise appearing through the coupling to the velocity field.
In the absence of chirality, the \emph{achiral} active force $\zeta $ leads to an effective layer tension \cite{Tapan1, Tapan2} which is stabilising when $\zeta <0$ and \emph{destabilises} the layered state when $\zeta >0$. 
However, \eqref{linstab2d} implies that the layered state is \emph{unstable} for \emph{any} value of $\zeta_c$ -- that is, chiral layered state \emph{do not exist} in two dimensions. To see this more clearly, we rewrite the dispersion relation implied by \eqref{linstab2d}:
\begin{equation}
\omega=-\frac{i}{\eta q ^2}\left( B\frac{q_z^2q_x^2}{ q ^2}-\zeta q_x^2-\zeta_cq_xq_z\right)=-\frac{i}{\eta}\left[\frac{ B}{4}\sin^2 (2\theta_q)-\zeta \sin^2 \theta_q-\frac{\zeta_c}{2}\sin (2\theta_q)\right],
\end{equation}
where $\theta_q$ is the angle between $\hat{\bf z}$ and the wavevector direction. It is then clear that the relaxation rate is \emph{negative}, implying an instability for \emph{any} sign and value of $\zeta_c$: for $\zeta_c>0$, this instability happens for $\theta_q\gtrsim 0$ while for $\zeta_c<0$, the relaxation rate is negative for $\theta_q\lesssim 0$.
Since the chiral layered state is \emph{generically} unstable at small wavenumbers, the coefficient of the $\mathcal{O}(q^2)$ term controls the stability at larger wavevectors and is required to obtain the lengthscale of the patterned state beyond the generic instability. 
We therefore expand the equation for $ u$ to $\mathcal{O}(q^2)$:
\begin{multline}
-i\omega  u=-\frac{1}{\eta q ^4}\{ Bq_z^2q_x^2-(\zeta q_x^2+\zeta_cq_xq_z)q^2\} u-\frac{1}{\eta}\left( Kq_x^2+2 B\chi_v\frac{q_z^3q_x}{ q ^2}-\chi_v\zeta_cq_z^2-\chi_v\zeta q_zq_x\right) u\\-\left[({\Gamma}_u B+\Lambda_1)q_z^2+\chi q_zq_x-\Lambda_2 q ^2\right] u-\frac{1}{\eta q ^4}(q_x^2\xi_{vz}-q_zq_x\xi_{vx}).
\end{multline}
The nonlinear equation of motion of the displacement field, explicitly writing the wavevector index, and retaining only the lowest order nonlinearities, is 
\begin{multline}
\partial_t u_q+\frac{i}{\eta k^2}\left[\left(\zeta_ck_zk_x+\zeta k_x^2-B\frac{k_z^2k_x^2}{k^2}\right)(q-k)_z-\left(\zeta_ck_z^2+\zeta k_zk_x-B\frac{k_z^3k_x}{k^2}\right)(q-k)_x\right]u_{k} u_{q-k}\\=-\left[\frac{1}{\eta q ^4}\{ Bq_z^2q_x^2-(\zeta q_x^2+\zeta_cq_xq_z)q^2\} +\frac{1}{\eta}\left( Kq_x^2+2 B\chi_v\frac{q_z^3q_x}{ q ^2}-\chi_v\zeta_cq_z^2-\chi_v\zeta q_zq_x\right) +\left\{({\Gamma}_u B+\Lambda_1)q_z^2+\chi q_zq_x-\Lambda_2 q ^2\right\}\right] u_q.
\end{multline}
The variance of the noise in this equation diverges at small $q$ as $1/q^2$. This nonlinear equation of motion needs to be solved, perhaps numerically, for obtaining the steady state in the presence of chirality and activity.

As with the Simha-Ramaswamy instability \cite{Aditi1} however, this instability can be made to acquire a finite threshold by bounding the system in the $y$ direction. Take $q_x=\pi/d$ i.e., consider a layer of width $d$ along the $y$ direction and consider a non-permeative state with $\zeta <0$ for simplicity. Also, assume that $\chi_v=0$. In this case, for $q_z\ll 1/d$,
\begin{equation}
\omega=-\frac{i}{\eta}\left[ B\left(\frac{q_zd}{\pi}\right)^2-\zeta -\zeta_c\left(\frac{q_zd}{\pi}\right)\right]
\end{equation}
which is stable at small $q_z$. Unlike the Simha-Ramaswamy instability of active nematics whose threshold vanishes as $1/d^2$, the threshold for this instability vanishes as $1/d$.


The generic, chiral instability \emph{is not} eliminated for non-permeative layers even when they are placed in contact with a substrate which acts as a momentum sink. The results for this case can be obtained by replacing $\eta\nabla^2$ in \eqref{linvybar} and \eqref{linvxbar} by the wavevector-independent friction $-{\Gamma}$ (and a corresponding non-conserving noise) and yields (without permeation)
\begin{equation}
\label{subsnoper}
-i\omega  u=-\frac{1}{{\Gamma}}\left( B\frac{q_z^2q_x^2}{q^2}-\zeta q_x^2-\zeta_cq_xq_z\right) u-\frac{q_x^2\xi_{vz}-q_zq_x{\xi}_{vx}}{\Gamma q ^2},
\end{equation}
which is clearly unstable at small wavevectors with a growth rate that now vanishes as $\sim q^2$. Note that in \eqref{subsnoper}, $\boldsymbol{\xi}_v({\bf r}, t)$ is a \emph{nonconserving} noise (since the momentum density is not a conserved quantity) whose correlator is $\langle{\boldsymbol{\xi}}_v({\bf r}, t){\boldsymbol{\xi}}_v({\bf r}', t')\rangle= 2{D}_v{\bsf I}\delta({\bf r}-{\bf r}')\delta(t-t')$. However, this active chiral instability can be prevented by \emph{permeation}. In this case, the equation of motion for $ u$ is 
\begin{equation}
-i\omega  u=-\frac{1}{{\Gamma}}\left( B\frac{q_z^2q_x^2}{q^2}-\zeta q_x^2-\zeta_cq_xq_z\right) u-\left[({\Gamma}_u B+\Lambda_1)q_z^2+\chi q_zq_x-\Lambda_2 q ^2\right] u-\frac{q_x^2\xi_{vz}-q_zq_x\xi_{vx}}{\Gamma q ^2}+{\xi}_u,
\end{equation}
which is not generically unstable (it may, however, be unstable for certain values of the active parameters).
When the active, chiral layered state with permeation is stable, the static structure factor of displacement fluctuations is 
\begin{equation}
\langle | u|^2\rangle=\frac{{D}_v(q_x^2/ q ^2)+{D}_u}{q_x^2[ B (q_z^2/ q ^2)-\zeta -\zeta _{c}(q_z/q_x)]+{\Gamma}[({\Gamma}_u B+\Lambda_1)q_z^2+\chi q_zq_x-\Lambda_2 q ^2]}.
\end{equation}
This static structure factor \emph{diverges} as $1/ q ^2$ along \emph{all} directions of wavevector space. This implies that this phase has quasi-long-range order in two dimensions. Moreover, in this case, it can be easily checked that there is \emph{no} relevant nonlinearity. Therefore, within a pure phase description, permeative two-dimensional chiral layered states on a substrate \emph{can} have quasi-long-range order.

A two-dimensional chiral layered state, with an \emph{in-plane} incompressibility constraint, at the \emph{interface} of two three-dimensional momentum conserved fluids is also generically unstable. The calculation for this case is similar to the one for the Stokesian momentum conserved with the only difference being the form of the mobility: instead of $1/\eta q ^2$, it is $1/2\eta| q |$ \cite{BroLen} and the dispersion relation is 
\begin{equation}
\omega=-\frac{i}{2\eta| q |}\left( B\frac{q_z^2q_x^2}{ q ^2}-\zeta q_x^2-\zeta_cq_xq_z\right)=-\frac{i| q |}{2\eta}\left[\frac{ B}{4}\sin^2 (2\theta_q)-\zeta \sin^2 \theta_q-\frac{\zeta_c}{2}\sin (2\theta_q)\right].
\end{equation}
While unlike a two-dimensional system in which the momentum is completely conserved in the plane, the growth rate of the instability vanishes linearly with the wavevector in this case, the instability cannot be stabilised by the permeative terms which appear at subleading order $\mathcal{O}( q ^2)$ in wavevectors.

The more realistic case with a \emph{three-dimensional} incompressibility constraint i.e., $\nabla\cdot{\bf v}=0$ where ${\bf v}\equiv(v_x,v_y, v_z)$ and $\nabla\equiv(\partial_x,\partial_y,\partial_z)$ are the three-dimensional velocity field and gradient operator respectively, (but where the pattern and the active and passive forces are still purely confined to the plane) is more interesting. The force balance equations in this case are
\begin{equation}
\label{linvybar2}
{\eta}{\nabla}^2{v}_x=\partial_x{ p}-\zeta_c\nabla^2 u\delta(y)-\chi_v\partial_z^2( B\partial_z^2 u- K\nabla^4 u)\delta(y)+{\xi}_{vx}
\end{equation}
\begin{equation}
{\eta}{\nabla}^2{v}_y=\partial_y{ p}+{\xi}_{vy},
\end{equation}
and
\begin{equation}
\label{linvxbar2}
{\eta}{\nabla}^2{v}_z=-( B\partial_z^2 u- K\nabla^4 u)\delta(y)+\partial_z{ p}+\zeta \nabla^2 u\delta(y)+\chi_v\partial_x\partial_z( B\partial_z^2 u- K\nabla^4 u)\delta(y)+{\xi}_{vz},
\end{equation}
where $\eta$ is the three-dimensional viscosity (we assume that the fluids are viscosity-matched) and $ p$ is the pressure that enforces the \emph{three-dimensional} incompressibility constraint. Using the three-dimensional transverse projector to solve for the velocity field and integrating it over all $q_x$, we obtain the dispersion relation for $ u$ to lowest order in wavevectors:
\begin{equation}
\omega=\frac{i}{4\eta| q |}\left[- B\frac{(2q_z^2q_x^2+q_z^4)}{ q ^2}+\zeta (q_z^2+2q_x^2)+\zeta_cq_xq_z\right]=\frac{i| q |}{4\eta}\left[- B\cos^2\theta_q(1+\sin^2\theta_q)+\zeta (1+\sin^2\theta_q)+\frac{1}{2}\zeta_c\sin(2\theta_q)\right]
\end{equation}
This is \emph{not} unstable for $\theta_q\to 0$. That is, the instability of the two-dimensional Stokesian layered state is eliminated when the layered state is formed at the \emph{interface} of two three-dimensional fluids with \emph{three-dimensional} momentum conservation {and}, crucially, three-dimensional fluid incompressibility.

Finally, if we consider a two-dimensional momentum-conserved system beyond the Stokes regime, we have to replace the $\eta\nabla^2$ on the R.H.S. of \eqref{linvybar} and \eqref{linvxbar} by inertia $-\rho_0\partial_t v_x$ and $-\rho_0\partial_t v_z$ respectively, where $\rho_0$ is the total mass density field (which is incompressible). This then yields a pair of sound waves with $\omega=\pm c(\theta_q) q $ 
\begin{equation}
c(\theta_q)=\sqrt{\frac{B\sin^2(2\theta_q)-4\zeta \sin^2\theta_q-2\zeta_c\sin(2\theta_q)}{4\rho_0}}.
\end{equation}
As expected, the sound speed turns imaginary for $\theta_q\lesssim 0$ when $\zeta_c<0$ and $\theta_q\gtrsim 0$ when $\zeta_c>0$.
\begin{figure}[t]
	\centering
	\includegraphics[width=7cm]{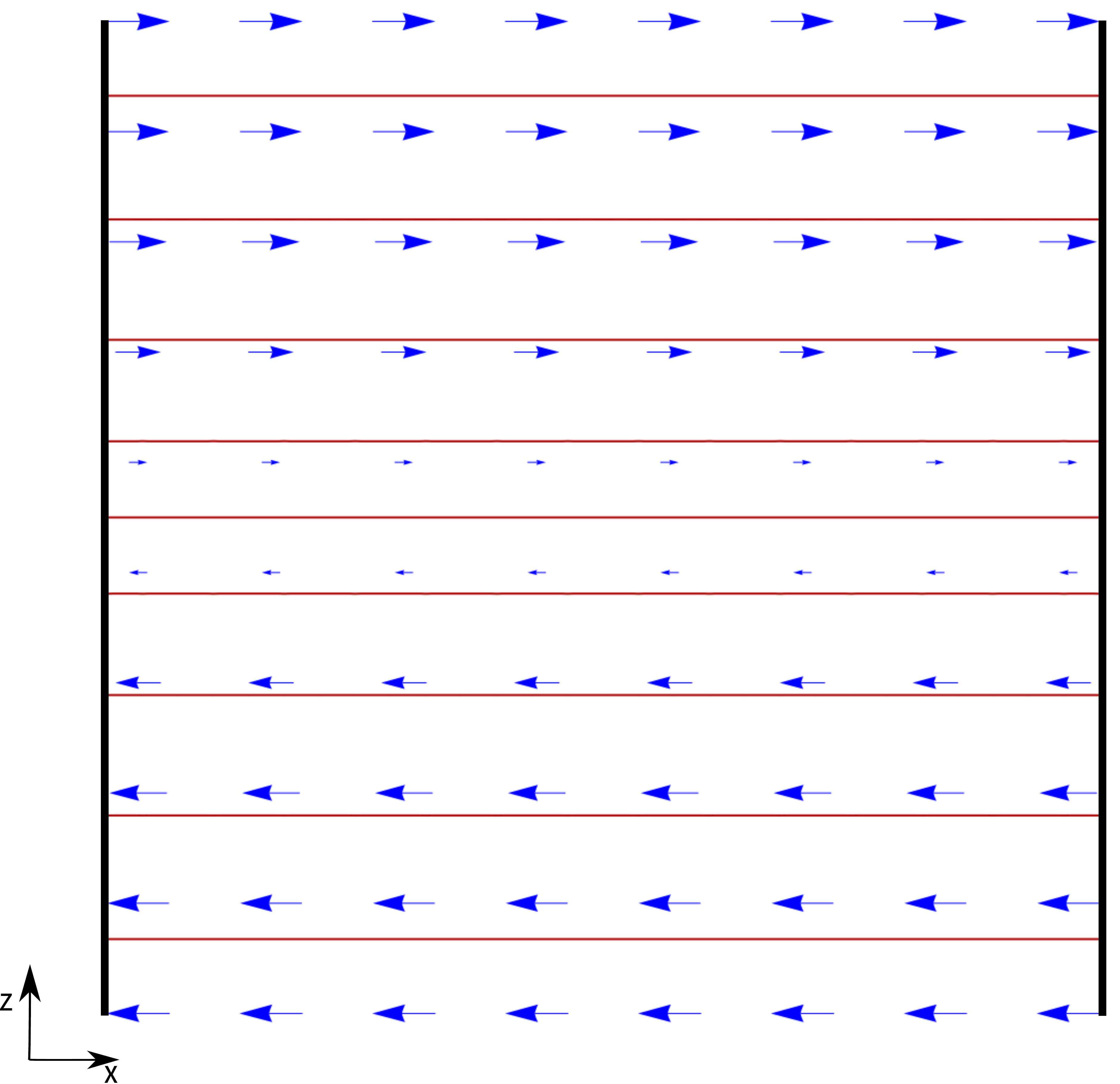}
	\caption{The flow field  due to the chiral, active force for a distortion of a two-dimensional layered state purely along the normal direction.}
	\label{fi3}
\end{figure}

It is known that the Simha-Ramaswamy instability of a \emph{polar}, inertial fluid can be eliminated for sufficiently high motility \cite{Rayan}. Can an array of \emph{motile} lines in a chiral active system \emph{also} escape the chiral generic instability? If we consider a layered state which is \emph{polar} in addition to being chiral, the lines will move in the direction of the polarisation field, which we take to be slaved to the normal to the layers. In this case, $\partial_t  u$ in \eqref{linubar} should be replaced by $\partial_t  u+\gamma[\partial_z u-(\nabla u)^2/2]$ \cite{ChenToner} where $\gamma$ is a phenomenological coefficient which carries information about the polarity. Considering only its linear part, the characteristic equation is then modified to 
\begin{equation}
\omega^2-\gamma\omega\cos\theta_q q -\frac{B\sin^2(2\theta_q)-4\zeta \sin^2\theta_q-2\zeta_c\sin(2\theta_q)}{4\rho_0} q ^2=0.
\end{equation}
This has the solution
\begin{equation}
\omega=\frac{1}{2}\left[\gamma\cos\theta_q\pm\sqrt{\frac{1}{\rho_0}}\sqrt{\gamma^2\cos^2\theta_q+B\sin^2(2\theta_q)-4\zeta \sin^2\theta_q-2\zeta_c\sin(2\theta_q)}\right].
\end{equation}
It is clear that when $\zeta <0$ and $\zeta_c$ is sufficiently small, the angular factor under the square root is \emph{not} negative for \emph{any} $\theta_q$. Indeed, it is not negative for small $\theta_q$, since $\gamma^2\cos^2\theta_q\approx\gamma^2\gg 0$, thus eliminating the generic instability. Therefore, inertia always suppresses the instability of \emph{polar} chiral layered states which would be present in \emph{Stokesian} fluids (for a strictly Stokesian fluid, the $\gamma$ term would only contribute at subleading order in wavenumbers).

\subsection{Linear theory of three-dimensional, chiral layered states}
\label{linchisec3d}
In this section we will examine the theory of linearised fluctuations about a layered state implied by \eqref{ueqn3d} and \eqref{vel3du} in Sec. \ref{3dconslayer} or equivalently, \eqref{ueqn3d2} and \eqref{vel3du2} in Sec. \ref{3dlaynoncons}. The linearised equations of motion are
\begin{equation}
\label{linu}
\partial_t u=v_z+ C_v\nabla^2(\partial_xv_y-\partial_yv_x)+\Lambda_1\partial_z^2 u-\Lambda_2\nabla^2 u+\Gamma_u(B\partial_z^2u-K\nabla^4u)+\xi_u,
\end{equation}
\begin{equation}
\label{linvx}
\eta\nabla^2v_x=\partial_x p+\zeta\partial_z\partial_x u+ z_c\partial_y\nabla^2u- C_v\partial_y\nabla^2(B\partial_z^2u-K\nabla^4u)+\xi_{vx},
\end{equation}
\begin{equation}
\label{linvy}
\eta\nabla^2v_y=\partial_y p+\zeta\partial_z\partial_y u- z_c\partial_x\nabla^2u+ C_v\partial_x\nabla^2(B\partial_z^2u-K\nabla^4u)+\xi_{vy}
\end{equation}
and
\begin{equation}
\label{linvz}
\eta\nabla^2v_z=-(B\partial_z^2u-K\nabla^4u)+\partial_z p+\zeta(\nabla^2+\partial^2_z) u+\xi_{vz}.
\end{equation}
The $ C_v$ coupling which is allowed even in \emph{passive} layered chiral systems, such as cholesterics, implies that the Laplacian of an in-plane vortical flow leads to motion of the layers. This is in contrast to \emph{achiral} layered systems, such as smetics and lamellar phases, in which in-plane flows do not lead to any displacement of layers within the linear theory. This implies that while the hydrodynamics of \emph{passive} smectic and cholesteric phases \emph{are} equivalent to the leading order, a distinction between them arises at a subleading  order. The active permeative terms in \eqref{linu} are both achiral and are, in principle, present even in active smectics \cite{Tapan1}. The only other effect of chirality arises through  the chiral active stress which, unlike in two dimensions, is nominally subdominant to the \emph{achiral} active stress. However, we will demonstrate in this section, and in greater detail, in the next one that this nevertheless has important qualitative consequences. Before embarking on that, we briefly comment on the leading order displacement field dynamics which is indistinguishable from an \emph{achiral} layered state. eliminating the pressure using the constraint of incompressiblity and eliminating the pressure, we obtain the Fourier-transformed equation of motion for the displacement field which, to the lowest order in wavenumbers is
\begin{equation}
-i\omega u=-\frac{1}{\eta q^2}\left[\left(B\frac{q_z^2q_\perp^2}{q^2}-\zeta q_\perp^2\right)u-q_z(q_x\xi_{vx}+q_y\xi_{vy})+q_\perp^2\xi_{vz}\right],
\end{equation}
where $q_\perp^2\equiv q_x^2+q_y^2$. The layered state is stable only when $\zeta<0$.
In this case, the static structure factor of displacement fluctuations is simply
\begin{equation}
\label{ssfmcons3d}
\langle |{u}|^2\rangle=\frac{{D}_v}{{\eta}[{B} q_z^2-\zeta q^2]},
\end{equation}
which diverges as $1/q^2$ in all directions of the wavevector space. Therefore, in three-dimensions, chiral, active layered states, in common with their achiral counterparts, but unlike \emph{passive} layered states, have long-range order.

We have shown that chirality doesn't modify the small wavenumber theory of layered states. However, it has a crucial effect nevertheless. From \eqref{linvx} and \eqref{linvy}, we see that the chiral active force (which is divergence-free) implies a vortical flow due to a curvature of the layers. To demonstrate this more clearly, we calculate the $z$ component of the vorticity field $\Omega_z\equiv\partial_x v_y-\partial_y v_x$ from \eqref{linvx} and \eqref{linvy}:
\begin{equation}
\eta\nabla^2\Omega_z=- z_c\nabla^2\nabla_\perp^2 u+ C_v\nabla^2\nabla_\perp^2(B\partial_z^2u-K\nabla^4u),
\end{equation}
with $\nabla_\perp^2\equiv\partial_x^2+\partial_y^2$, where, as expected, only the chiral active force and the chiral passive coupling to the displacement field appear. The passive coupling is subdominant to the active one and to lowest order,
\begin{equation}
\label{vortsim}
\Omega_z\sim-\frac{ z_c}{\eta}\nabla_\perp^2 u.
\end{equation}
As advertised earlier, this implies that a curvature of the layers lead to in-plane vortical flows. This is a unique feature of chiral and active layered states.

\section{Beyond linear theory: Spontaneous vortex lattice states}
\label{beyondlin}
\begin{figure}[t]
	\centering
	\includegraphics[width=15cm]{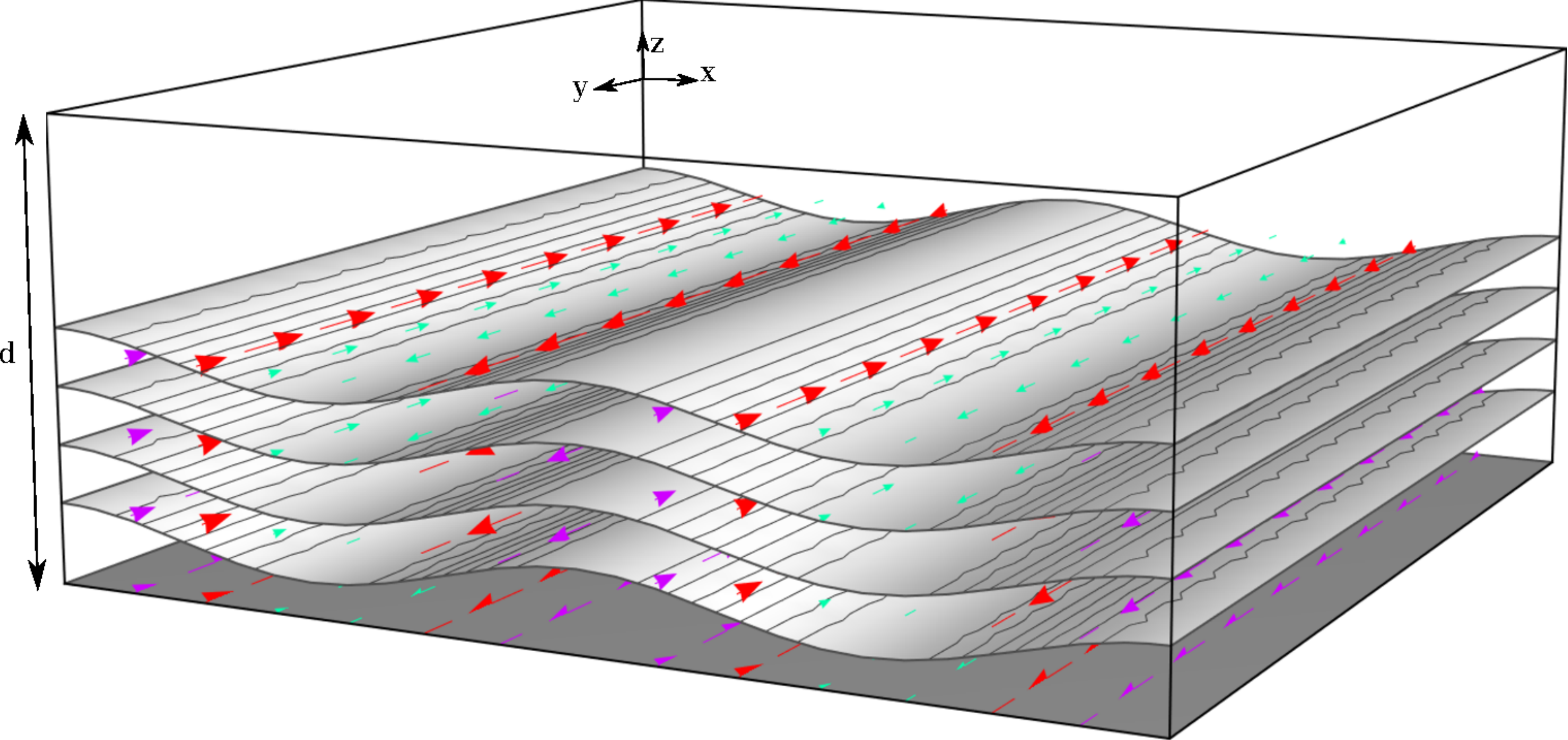}
	\caption{The active chiral flow field  due to one dimensional undulations in three dimensional layered states}
	\label{fi2}
\end{figure}

In the last section we demonstrated that a curvature of layers must, inevitably lead to in-plane vortical flows in chiral, active layered states in three dimensions. In this section, considering non-permeative layered states for simplicity, we will demonstrate that the \emph{achiral} active force can lead to a spontaneous formation of state with periodic layer undulations in the $x-y$ plane. Then, the effect of the \emph{chiral} active force would be to create an in-plane vortex lattice state with \emph{counter-rotating} fluid vortices. 
As discussed in the last section, the layered state in an infinite system is unstable for $\zeta>0$. Further, we demonstrated in Sec. \ref{Eqastrestr} that a state with an active stress $\zeta$ is equivalent to a passive layered system under an \emph{external} stress $\sigma_0=\zeta$. Therefore, the threshold-free instability of an \emph{active} layered state for $\zeta>0$ maps onto the Helfrich-Hurault instability of an externally stressed layered state, the threshold for which vanishes with the sample size. In this case, when the layered state is destabilised, a square undulated pattern (an egg-crate-like structure) is realised. 
Based on the equivalence of external stress and the achiral active stress, we therefore argue that a similar pattern should be realised beyond the active instability as well. The wavelength of the undulations have to be $q_p\propto \sqrt{\zeta/K}$. Therefore, a displacement field just beyond the instability of the form
\begin{equation}
u=u_0\cos q_px\cos q_py
\end{equation}
leads to an in-plane vorticity pattern in a chiral active system from \eqref{vortsim} of the form
\begin{equation}
\Omega_z\propto \frac{ z_c q_p^2 u_0}{\eta}\cos [q_p(x+y)].
\end{equation}
This is the vortex lattice depicted in the main text. Crucially, this vortex lattice state arises spontaneously from the active instability. In the main text, we discuss how to control the structure of the vortex-lattice using the mapping between the active and the external stress.
The vortex lattice discussed in the main text results from a distortion of the layered state in a square, egg-crate-like pattern. However, other distortions are possible; for instance, a one-dimensional undulatory pattern which we take to be along $\hat{x}$. In this case, the vortices will get infinitely stretched along the $y$ direction. We display the flow field generated in that case in Fig. \ref{fi2}.

